\documentclass[a4paper,final]{jfm}

\usepackage{graphicx}
\usepackage{newtxtext}
\usepackage{newtxmath}
\usepackage{natbib}
\usepackage{hyperref}
\hypersetup{
    colorlinks = true,
    urlcolor   = blue,
    citecolor  = black,
}

\usepackage{bm}
\usepackage{multicol}
\usepackage{multirow}
\usepackage{caption,subcaption}

\makeatletter
\renewcommand{\maketag@@@}[1]{\hbox{\m@th\normalsize\normalfont#1}}
\makeatother

\newtheorem{theorem}{Theorem}

\newtheorem{remark}{Remark}[subsection]
\newcommand{\RomanNumeralCaps}[1]
\linenumbers

\newcommand{{\gminus}}{f}

\title{Singularity of the axisymmetric stagnation-point-like solution within a cylinder of the 3D Euler incompressible fluid equations}

\author{Yinshen Xu\aff{1}\corresp{ 
\email{xuyinshen@163.com, miguel.bustamante@ucd.ie}}, Miguel D. Bustamante\aff{1} }

\affiliation{\aff{1}School of Mathematics and Statistics, University College Dublin, Belfield, Dublin 4, Ireland}

\begin{document}

\maketitle

\begin{abstract}
In this paper we investigate analytically the formation of finite time singularities in the three dimensional incompressible Euler equations under the model of Gibbon, Fokas, and Doering for vorticity stretching within a bounded cylindrical domain and under axisymmetric conditions. We derive explicit Lagrangian solutions for the vorticity, its stretching rate, fluid pathlines, and velocity components by exploiting constants of motion associated with the field dependent infinitesimal symmetries of the system. The central finding is that the existence and nature of a finite time singularity are determined exclusively by the local geometric structure of the initial vortex stretching rate near its global minimum. Whether a singularity forms depends on how flat this profile is at the minimum. Flatter profiles delay the blowup and sufficient flatness can suppress it entirely. For power law behavior near the minimum, critical thresholds for the exponent are identified which separate regular solutions from those that develop a finite time singularity. These thresholds differ depending on whether the singularity occurs at the center of the cylinder or on a ring away from the center, with minima at the center requiring higher flatness to avoid blowup. This work provides a rigorous analytical framework that elucidates how the local geometric structure of the initial conditions governs the potential for singularity formation in 3D fluid flows, offering fundamental insights into the interplay between symmetry, initial data, and the development of extreme events in idealized turbulence.
\end{abstract}

\section{Introduction}\label{introduction}

In fluid mechanics, the conservation of momentum for an inviscid, adiabatic fluid is elegantly expressed by Euler’s equation—a cornerstone of the field whose significance cannot be overstated. Since its formulation in 1761 by Leonhard Euler, this equation has had a profound and lasting impact on mathematics, physics, and engineering \citep{euler1761principia}. 
We define a \textbf{3D incompressible Euler fluid} by its velocity and pressure fields. Let  
$\bm{u}: \Omega \times [0,T) \rightarrow \mathbb{R}^3$ 
be a vector field and  
$p: \Omega \times [0,T) \rightarrow \mathbb{R}$
be a scalar field, where $\Omega \subseteq \mathbb{R}^3$ denotes the spatial domain and $[0,T) \subseteq [0,\infty)$ represents the time domain, with $T>0$. Because we will consider classical solutions and explicit formulas where the singularity time is computable in terms of the initial condition, we will assume these fields to be of class $C^k$ piecewise in terms of spatial derivatives, where $k$ is appropriately chosen. If
\begin{equation}\label{euler}
    \frac{\partial \bm{u}}{\partial t} + (\bm{u} \cdot \nabla) \bm{u} = -\nabla p\,, \qquad \nabla \cdot \bm{u} = 0\,, \qquad \text{for \quad all} \qquad \bm{x} \in \Omega\,, \quad  t \in [0,T)\,,
\end{equation}  
where $\nabla$ is the gradient operator, then $\bm{u}$ and $p$ represent the velocity and pressure fields of a {3D incompressible Euler fluid}, and equation~\eqref{euler} is referred to as the \emph{3D Euler equation for an incompressible fluid} (the incompressibility condition is $\nabla \cdot \bm{u} = 0$).

 {The vorticity, defined as $\bm{\omega} := \nabla \times \bm{u}$, satisfies the Helmholtz vorticity equation obtained by taking the curl of the Euler equations \eqref{euler}:}
\begin{equation}
\label{helmholtz}  
    \frac{\partial \bm{\omega}}{\partial t} + \left(\bm{u}\cdot\nabla\right)\bm{\omega} - \left(\bm{\omega}\cdot\nabla\right)\bm{u} = 0\,, \qquad \nabla \cdot \bm{\omega} = 0\,, \qquad \text{for \quad all} \qquad \bm{x} \in \Omega\,, \quad  t \in [0,T)\,,
\end{equation}                                              
From its definition as a curl, it is obvious that $\nabla\cdot\bm{\omega}=0$.

One of the most intriguing topics in mathematical fluid dynamics is whether a finite-time singularity emerges in the solution of the 3D Euler fluid equations \eqref{euler}.  Many numerical experiments have revealed the existence of a finite-time singularity, see for example \cite{gibbon2008three,luo2014potentially,choi2017finite,larios2018computational,hou2023potentially}. A milestone of the investigation of the regularity of the solution to the 3D Euler equations is the so-called  {Beale-Kato-Majda } theorem 
\citep{beale1984remarks} which points out that the convergence of the time integral of the supremum norm (over the spatial domain) of vorticity is a necessary and sufficient condition for 3D Euler fluids to maintain regularity. Subsequently, 
\cite{constantin1996geometric} proved that the finite norm of the velocity field along with vorticity direction vectors of class $C^1$ constitute a necessary condition to keep 3D Euler flows regular within a finite time. 
 {Over the years, considerable research has been conducted on the regularity properties and local well-posedness of the 3D Euler equations; see the recent work by \cite{drivas2023singularity}. While general approaches provide broad theoretical insights, they often do not yield detailed information about the structure or dynamics of potential singularities. In contrast, imposing additional structure—such as axisymmetry \citep{elgindi2020finite}, self-similarity \citep{martinez2024evidence}, or octahedral symmetries \citep{elgindi2021incompressible}—can significantly simplify the analysis and make it possible to probe singularity formation with greater precision. }

When we further impose a simplifying ansatz, the properties of the Euler flows may even be manifested by explicit equations and solutions. One such ansatz is the stagnation-point-type flow proposed by Gibbon, Fokas and Doering \citep{gibbon1999dynamically}, featuring the vortex stretching rate $\gamma(x,y,t)$ (see also \cite{stuart1988Nonlinear} and \cite{lundgren1982strained}):
\begin{equation}\label{gibbon}
    \bm{u}(x,y,z,t) = \left(u_x(x,y,t), u_y(x,y,t), z\gamma(x,y,t)\right),
\end{equation}
where
\begin{align}
    (x,y,z,t)\in\Omega_2 \times \mathbb{R} \times[0,T)\,, \quad \Omega_2 \subset \mathbb{R}^2.
\end{align}
In the following, we will name this {``Gibbon-Fokas-Doering vorticity-stretching model'', or simply ``Gibbon-Fokas-Doering model''}. Evidently, system \eqref{gibbon} contains infinite energy, suggesting that if a finite-time singularity  occurs then it will do so not at a point but on a one-dimensional manifold of the spatial domain. Physically speaking, this singularity with infinite energy demonstrates a change in some topological structure or the breakdown of the stagnation-type ansatz in \eqref{gibbon}. In what follows we list previous studies that, to the best of our knowledge, are relevant to {Gibbon-Fokas-Doering}  stagnation-type model.

As early as 1989, Childress, Ierley, Spiegel and Young exhibited blowup in their numerical analysis of two-dimensional stagnation flow \citep{childress1989blow}. As for the three-dimensional case, equation \eqref{gibbon}, the first numerical experiment was implemented by \cite{ohkitani2000numerical}, discovering the finite-time blowup for specific initial conditions and finding the approximate inverse proportional relation $\displaystyle\max_{(x,y)\in\Omega_2\subseteq\mathbb{R}^2}|\gamma(x,y,t)|\propto(T_*-t)^{-c}$ with $c\approx1$, where $T_*$ denotes the singularity time and $\Omega_2 = \mathbb{T}^2$, the $2$-torus. Shortly afterwards, an analytic Lagrangian solution for the stretching rate $\gamma(x,y,t)$ was given by  \cite{constantin2000euler} in the same case $\Omega_2 = \mathbb{T}^2$. In the limiting case $\Omega_2 = \mathbb{R}^2$ of an infinite domain, Gibbon, Moore and Stuart obtained, in the case of zero plane vorticity $\omega$, explicit axisymmetric solutions for $\gamma$ and pathlines that indeed have finite-time singularity \citep{gibbon2003exact}. All these solutions find their generalization in the work of  \cite{bustamante2022role} which introduces a rich Lie-algebra structure of infinitesimal contemporaneous symmetries, based on which a general Lagrangian solution of stretching rate and vorticity are attained analytically. Also, to model the corrections that would represent a finite-energy solution, Mulungye, Lucas and Bustamante put forward a pressure field modified on the second derivative in the vertical direction, modeled in terms of a free parameter, while Lagrangian solutions are still obtained analytically  \citep{mulungye2015symmetry}. Detailed analysis of the existence and asymptotic behaviors of the singular solutions to the 3D Euler fluid equation under this modified pressure field are illustrated, where the asymptotic form of $\gamma$ is certainly derived as $\displaystyle\max_{(x,y)\in\Omega_2\subset\mathbb{R}^2}|\gamma(x,y,t)|\propto(T_*-t)^{-1}$ \citep{bustamante2022role, mulungye2015symmetry, mulungye2016atypical}.

Of note is the numerical experiment by \cite{luo2014potentially} of an axisymmetric solution of 3D incompressible Euler equation inside a cylinder (with a slip boundary condition), which disclosed a ring-like singularity located at the cylindrical boundary. Note also that a similar ring-like grow-up (in a domain without boundary $\Omega = \mathbb{T}^2 \times \mathbb{R}$) was observed by Ohkitani and Gibbon in their numerical simulation of {Gibbon-Fokas-Doering} model \cite[Initial condition 2]{ohkitani2000numerical}.  {However, as we will explain at the end of section \ref{conclusion}, that configuration does not lead to a finite-time singularity  because the local profile of the vorticity stretching rate is `flat' enough at the ring.}
Therefore, the axisymmetric solution of {Gibbon-Fokas-Doering} vorticity-stretching model for a domain with boundary is worthy of an investigation.  {While it is known that finite-time singularity formation in this model depends on the initial stretching rate through an integral condition \citep{constantin2000euler, bustamante2022role}, the precise nature of this dependence—whether governed by local or global features, algebraic or geometric structure—has remained unclear. } 
 {In this paper, we show that finite-time singularity formation in this model is governed entirely by the local power-law exponent of the initial stretching rate at its minimum: blowup occurs below a critical flatness threshold, and global regularity prevails above it.}

 {Although the stagnation-point-like solution of \cite{gibbon1999dynamically}  possesses infinite energy, it captures the essential local dynamics of a vortex tube embedded within a larger, finite-energy flow. The singularity criteria derived here are purely local: they depend only on the geometric structure of the initial stretching rate in a neighborhood of its minimum. This implies that our results can serve as a diagnostic tool to assess whether a localized vortex tube in a realistic flow is prone to finite-time breakdown—after which the tube effectively ``disappears'' from the perspective of this reduced description. }

The rest of the paper is arranged as follows. In Section \ref{Euler}, the Lagrangian solutions for the vorticity, its stretching rate and pathlines are derived explicitly, including a formula for the singularity time $T_*$. Moreover, new explicit Lagrangian solutions for the velocity field components are obtained, thanks to the axisymmetry assumption. Next, Section \ref{singular} discusses the existence of a finite-time singularity and the asymptotic behavior of the solutions:  subsection \ref{parabola} gives a couple of examples of ``parabolic'' initial stretching rate to exhibit how singularity behaves and to provide new explicit Eulerian solutions for the velocity field, the vorticity, its stretching rate and the pathlines;  subsection \ref{continuous} investigates the general case of initial stretching rate whose profile approaches the minimum value as a power law, with (possibly) kinks at the positions of the minimum, and how these power-law exponents determine singular behaviors. Finally, conclusions are given in Section \ref{conclusion}.

\section{Axisymmetric Solution to the 3D Euler Incompressible Fluid Equations within {Gibbon-Fokas-Doering}  Vorticity-Stretching Model Ansatz}\label{Euler}

In this section, we consider the three-dimensional incompressible Euler fluid under axisymmetric conditions{, in the context of the ansatz from} \cite{gibbon1999dynamically}, and derive explicit Lagrangian solutions {(i.e., solutions along the pathlines)} for the vorticity, {vortex stretching rate, and the pathlines themselves, in terms of the initial conditions for the vorticity and the vortex stretching rate. A new feature arising from the axisymmetry condition is that implicit and explicit Lagrangian solutions for the velocity field components can be found. Moreover, Eulerian solutions for all these fields can be found for certain initial conditions.} {These} solutions reveal the potential formation of singularities associated with the initial vortex stretching rate, which motivates the subsequent {more detailed} analysis of singularity behavior in Section \ref{singular}.

{To take advantage of the axisymmetric conditions, we further restrict the geometry of the fluid to} a cylindrical region. Specifically:

\begin{itemize}
    \item[a)] Spatial Domain: The spatial coordinates are defined as $ \bm{X} = (r, \theta, z) \in \Omega_C = [0, R] \times [0, 2\pi) \times \mathbb{R} $, indicating that the fluid is bounded by a cylindrical region of radius $ R $ and extends infinitely along the vertical $ z $-axis. For convenience, we adopt cylindrical coordinates throughout the discussion, where $ \hat{r} $ denotes the radial direction centered at the cylinder’s axis ($ r = 0 $), $ \hat{\theta} $ represents the azimuthal direction oriented counterclockwise around the central axis, and $ \hat{z} $ is the vertical direction;
    
    \item[b)] Axisymmetry: The fluid is axisymmetric across the entire domain, meaning that both the velocity field $ \bm{u} $ and the pressure field $ p $ are independent of the azimuthal coordinate $ \theta $:
$\displaystyle    \frac{\partial \bm{u}}{\partial \theta} = \bm{0}, \quad \frac{\partial p}{\partial \theta} = 0 $, on $\Omega_C$ and at all times $t\in [0,T)$. 
\end{itemize}

\subsection{{Gibbon-Fokas-Doering} Stagnation-point-like Solution: Axisymmetric Solution within a Cylinder for an Inviscid Incompressible Fluid}\label{Gibbon}

 Gibbon, Fokas, and Doering proposed a fluid model in \citep{gibbon1999dynamically} to describe the dynamics of vortical structures arising from vortex interactions. This model has been shown to admit singular Lagrangian solutions under \textit{x}\textit{y}-periodic boundary conditions. In this work, we extend  {that model} to investigate the formation of singularities in solutions of the three-dimensional (3D) incompressible Euler equations under axisymmetric conditions within a cylindrical domain with one lid. 

We consider an inviscid, incompressible 3D fluid confined within a cylindrical spatial domain $\Omega_C = [0,R] \times [0,2\pi) \times \mathbb{R}$, where the velocity field is assumed to take the following form:
\begin{equation}\label{velo}
    \bm{u}(r,z,t) = u_r(r,t)\hat{r} + u_{\theta}(r,t)\hat{\theta} + z\gamma(r,t)\hat{z}, \quad (r,\theta,z,t) \in \Omega_C \times [0,T),
\end{equation}
with $T > 0$ denoting the maximal time interval over which the solution remains regular.  {We impose that $u_r(r,t),u_{\theta}(r,t),\gamma(r,t)$ be continuous on $[0,R]\times[0,T)$ and piecewise in $C_r^2C^1_t\left([0,R]\times[0,T)\right)$, which  denotes the space of functions with continuous second-order partial derivatives with respect to $r$ and continuous first-order partial derivatives with respect to $t$. This prepares us to search for strong solutions of the Euler equation  \eqref{euler} and the Helmholtz vorticity equation \eqref{helmholtz}.} The velocity field exhibits a stagnation plane at $z=0$, with radial and azimuthal components symmetric about this plane, while the vertical component is {antisymmetric}. Consequently, it suffices to analyze the fluid behavior in the subregion where $z \geq 0$.

To account for the cylindrical {boundary}, we impose the no-flow boundary condition at $r=R$:
\begin{equation*}
    u_r(r=R, t) = 0,
\end{equation*}
which ensures that the radial velocity vanishes at the cylindrical wall. No explicit boundary conditions are imposed on the azimuthal or vertical components of the velocity field {at the $r=R$ boundary}.

Given the velocity field defined above, the corresponding vorticity field $\bm{\omega} = \nabla \times \bm{u}$ is computed as:
\begin{equation}\label{vorti}
    \bm{\omega}(r,z,t) = - z\frac{\partial \gamma}{\partial r}(r,t)\hat{\theta} + \omega(r,t)\hat{z},
\end{equation}
where the axial component $\omega(r,t) := \bm{\omega} \cdot \hat{z}$ represents the vorticity restricted to the plane $z=0$, referred to as the \textbf{plane vorticity}. It is explicitly given by:
\begin{equation}\label{omegatout}
    \omega(r,t) = \frac{\partial u_\theta}{\partial r}(r,t) + \frac{u_\theta(r,t)}{r}.
\end{equation}
Moreover, the incompressibility condition $\nabla \cdot \bm{u} = 0$ leads to the relation:
\begin{equation}\label{gammatour}
    \gamma(r,t) = - \frac{\partial u_r}{\partial r}(r,t) - \frac{u_r(r,t)}{r}.
\end{equation}

\begin{remark}
The term ``{vorticity} stretching rate'' {(stretching rate, for short)} for $\gamma$ in relation to the plane vorticity $\omega$ is justified by the following observation. By considering the two-dimensional restriction of the   {Gibbon-Fokas-Doering model} on the plane $z = 0$ from equation \eqref{velo}, we can define the three-dimensional vorticity as $\omega \hat{z}$ and substitute it into the Helmholtz vorticity equation \eqref{vorti}, resulting in
\begin{equation*}
     {\left(\frac{D\omega}{Dt} =\right) \frac{\partial \omega}{\partial t} + u_r \frac{\partial \omega}{\partial r}= } \gamma \omega,
\end{equation*}
where  {$\frac{D}{Dt}:= \frac{\partial}{\partial t} + u_r \frac{\partial}{\partial r} + u_z \frac{\partial}{\partial z}$} denotes the material  {time} derivative. In this context, $\omega$ represents the vorticity of the reduced two-dimensional fluid. This equation shows that the  {material} rate of change of $\omega$ is directly proportional to $\omega$ itself, with proportionality coefficient $\gamma$. 
Hence, the quantity $\gamma$ in {the Gibbon-Fokas-Doering}  model \eqref{velo} is referred to as the stretching rate.
\end{remark}

It is evident from equations \eqref{gammatour} and \eqref{omegatout} that,  {at a given time $t$,} they represent linear ordinary differential equations in $u_r$ and $u_\theta$, respectively. Solving these yields explicit expressions for the radial and azimuthal velocity components:
\begin{equation}
    \label{urtogamma}
u_r(r,t) = -\frac{1}{r} \int_0^r \gamma(r',t) r' \, \mathrm{d}r', \quad u_\theta(r,t) = \frac{1}{r} \int_0^r \omega(r',t) r' \, \mathrm{d}r'.
\end{equation}
{The velocity field is the centerpiece} of the Eulerian description of a fluid,  {a role that the pathlines play in the Lagrangian description.}  Combining equation \eqref{urtogamma} with the velocity form given in equation \eqref{velo}, we deduce that the Eulerian velocity field can be fully expressed in terms of the plane vorticity $\omega$, its stretching rate $\gamma$, and the vertical coordinate $z$. Consequently, the triplet $(\omega, \gamma, z)$ captures all information necessary for describing the fluid dynamics within the Eulerian framework.

 {From the requirement of physical consistency of axisymmetric flows in cylindrical coordinates, the velocity must be single-valued at the axis $r=0$, which implies
\begin{equation*}
    u_r(r=0, t) = u_\theta(r=0, t) = 0, \quad t \in [0,T)\,,
\end{equation*}
which we impose as boundary conditions from the outset and is consistent with equation \eqref{urtogamma} when taking the limit $r\rightarrow0$, thanks to the continuity assumptions on $\gamma$ and $\omega$. These conditions eliminate the coordinate singularity at $r=0$ and are standard for physically admissible axisymmetric Euler flows, see \citep{batchelor2000introduction}.}

Furthermore, due to the no-flow boundary condition $u_r(r=R, t) = 0$, and the relation between $u_r$ and $\gamma$ in equation \eqref{urtogamma}, we deduce that
\begin{equation*}
    \langle \gamma(\cdot,t) \rangle = 0, \quad t \in [0,T)\,,
\end{equation*}
where $\langle \cdot \rangle$ denotes the volume-averaged integral over the cylindrical domain:
\begin{equation}
\label{eq:average}
    \langle f(\cdot,t) \rangle := \frac{1}{\pi R^2} \int_0^{2\pi} \int_0^R f(r,t) r \, \mathrm{d}r \, \mathrm{d}\theta.
\end{equation}

Substituting the velocity and vorticity fields defined in equations \eqref{velo} and \eqref{vorti} into the 3D Euler equation \eqref{euler} and the Helmholtz vorticity equation \eqref{helmholtz}, we derive the evolution equations for $\gamma$ and $\omega$:
\begin{equation}
    \label{motion}
\begin{cases}
\displaystyle \frac{\partial \gamma}{\partial t} + u_r \frac{\partial \gamma}{\partial r} + \gamma^2 = 2\left\langle \gamma^2 \right\rangle, \\
\\
\displaystyle \frac{\partial \omega}{\partial t} + u_r \frac{\partial \omega}{\partial r} - \gamma \omega = 0.
\end{cases}
\end{equation}

 {It is worth remarking that the term $2\langle \gamma^2 \rangle$ in the evolution equation for $\gamma$ is a nonlocal term that turns the system  \eqref{gammatour}, \eqref{motion} into a system of partial integro-differential equations. This nonlocal term originates from the pressure field's second derivative with respect to the $z$ coordinate. Because the vertical velocity is linear in $z$, consistency of the Euler equations in all three directions requires that the combination $\partial_t \gamma + u_r \partial_r \gamma + \gamma^2$ be spatially uniform (i.e., independent of $r$). This uniform value is then determined by taking the volume average over the cross-section. Using the fact that $\langle \gamma \rangle = 0$—a consequence of the incompressibility condition together with the no-flow boundary condition at $r=R$—the advective term averages to $\langle \gamma^2 \rangle$, leaving precisely $2\langle \gamma^2 \rangle$. See \citep{ohkitani2000numerical} for a related derivation.}

{We note that the system of  equations} \eqref{omegatout}, \eqref{gammatour}, \eqref{motion} is invariant under the following scale transformation:
\begin{equation}
    \label{invariant}
u_r \mapsto k u_r(kr, k^2 t), \quad u_\theta \mapsto k u_\theta(kr, k^2 t), \quad \omega \mapsto k^2 \omega(kr, k^2 t), \quad \gamma \mapsto k^2 \gamma(kr, k^2 t),
\end{equation}
for any positive constant $k$ and arbitrary fields $(u_r, u_\theta, \omega, \gamma)$. 
 {Under this transformation, a solution defined on the cylindrical domain $\Omega_C = [0,R] \times [0,2\pi) \times \mathbb{R}$ with no-flow boundary condition $u_r(R,t) = 0$ is mapped to a new solution on the rescaled domain $\Omega_C^{(k)} = \left[0, \frac{R}{k}\right] \times [0,2\pi) \times \mathbb{R}$, satisfying the corresponding boundary condition
\[
u_r^{(k)}(R/k, t) = k u_r\left(k \cdot \frac{R}{k}, k^2 t\right) = ku_r(R, k^2 t) = 0.
\]
Thus, the scaling symmetry preserves the form of the problem, including the no-flow condition, provided the cylinder radius is rescaled accordingly. This property allows us to generate an infinite family of solutions from a single reference solution by varying $k > 0$.}

\subsection{Solutions for Plane Vorticity, Stretching Rate, Pathlines, and Velocity Field}\label{solution}

Analytical solutions for the stretching rate and plane vorticity can be derived by employing infinitesimal contemporaneous symmetries and associated constants of motion. This approach was first introduced by \cite{bustamante2022role}, where constants of motion are defined as conserved quantities in the Lagrangian framework, which plays a central role in the present study. Explicitly, under the context of axisymmetry, a scalar function $C(r,z,t)$ is a constant of motion if it obeys the conservation law
\begin{equation}
\label{constant}
     {\frac{DC}{Dt} =} \frac{\partial C}{\partial t} + u_r \frac{\partial C}{\partial r} + u_z \frac{\partial C}{\partial z} = 0, \qquad \text{for \quad all} \quad r \in [0,R], \quad z\in \mathbb{R}, \quad t \in [0,T),
\end{equation}
where $ u_r $ and $u_z$ denote respectively the radial and vertical components of the velocity field.

 {In this subsection we provide a summary of useful results obtained by \cite{bustamante2022role}, who carried out a systematic search of infinitesimal contemporaneous symmetries and constants of motion for Gibbon-type vortex stretching models. In that work, the time evolution of the system is encoded in a single scalar function $A(t)$, which satisfies the equation
\begin{equation}
\label{eq:Awithgamma}
    \ddot{A}(t) + 2\left\langle \gamma^2 \right\rangle A(t) = 0,  \quad A(0) = 1,\quad \dot{A}(0) = 0, \quad t\in [0,T),
\end{equation}
where $ \langle \cdot \rangle $ denotes the volume average over the cylindrical domain (see equation \eqref{eq:average}). As we will see below, the definition \eqref{eq:AtoS} and the nonlinear ordinary differential equation \eqref{ode} determine $A(t)$ by quadrature. However, it is useful to think of equation \eqref{eq:Awithgamma} as a linear ordinary differential equation for $A(t)$.
This equation establishes a direct relation between the pressure's curvature $p_{zz}(t):=2\left\langle\gamma^2\right\rangle$ and the function $A(t)$: for any continuous $p_{zz}(t)$, the initial value problem for \eqref{eq:Awithgamma} admits a unique {twice continuously differentiable} solution $A(t)$ by the Picard–Lindel\"{o}f theorem.  
{Moreover, from our short-time continuity assumption for $\gamma$, there exists a local time interval $[0,t_*)$ such that  $A(t)>0$ for any $t\in[0, t_*)$.} We introduce now the function $S(t)$ {for $t\in[0, t_*)$} via 
\begin{align}
\label{eq:AtoS}
    S(t) := \int_0^t\frac{1}{A(t')^2}dt',
\end{align}
{with initial data $S(0) = 0, \,\dot{S}(0) = 1$ (and also  $\ddot{S}(0) = 0$), calculated directly from the initial conditions of $A(t)$ in equation \eqref{eq:Awithgamma}.} From this function one can construct several conserved quantities along fluid particle trajectories.  For the purpose of our analysis, we adopt three of the constants of motion derived in \cite{bustamante2022role} (see the paragraphs below equations (3.15) and (3.16) there): 
\begin{equation}
\label{constantofmotion}
    \begin{aligned}
    C_1(r,t) &= -\frac{1}{2}\omega(r,t)\dot{S}(t)^{-\frac{3}{2}}\left[\ddot{S}(t) + 2\gamma(r,t)\dot{S}(t)\right], \\
    C_2(r,t) &= \frac{2\dot{S}(t)^2}{\ddot{S}(t) + 2\gamma(r,t)\dot{S}(t)} - S(t), \\
    C_3(r,z,t) &= -\frac{1}{2} z \dot{S}(t)^{-\frac{3}{2}}\left[\ddot{S}(t) + 2\gamma(r,t)\dot{S}(t)\right].
\end{aligned}
\end{equation}
These quantities remain invariant along Lagrangian trajectories: $\frac{DC_j}{Dt} = 0$ for $j=1,2,3$, as can be verified from \eqref{motion} after tedious but straightforward calculations. Consequently,
\begin{align}
\label{eq:Cpath}
     C_1(r(r_0,t), t) = C_1(r_0, 0), \quad  C_2(r(r_0,t), t) = C_2(r_0, 0),\quad  C_3(r(r_0,t),z(r_0,z_0,t), t) = C_3(r_0,z_0, 0),   
\end{align}
where {$r(r_0,t)$ denotes the solution to $\mathrm{d}r/\mathrm{d}t = u_r(r,t)$ with} $ r(r_0,0) = r_0$, and $r_0$ denotes the initial radial position, whereas $z(r_0,z_0,t)$ denotes the solution to $\mathrm{d}z/\mathrm{d}t = u_z(r,z,t)$ with $ z(r_0,z_0,0) = z_0$, and $z_0$ denotes the initial vertical position.}

 {
\begin{remark}
    Among the conserved quantities derived in \cite{bustamante2022role}, the three constants {of motion} $C_1, C_2, C_3$ from equation (\ref{constantofmotion}) are particularly useful for our analysis: $C_1$ and $C_2$ allow us to reconstruct the Lagrangian evolution of $\gamma$ and $\omega$, while $C_3$ determines the vertical pathline $z(t)$. Although these constants {of motion} do not directly correspond to classical physical invariants such as energy or momentum, they encode the geometric structure of the flow along particle trajectories and are essential for constructing the Lagrangian solution.
\end{remark}
}
 {From the invariant relations \eqref{constantofmotion}--\eqref{eq:Cpath}}, the Lagrangian solutions for the stretching rate $ \gamma $ and the plane vorticity $ \omega $ are obtained directly.  {Specifically, the constancy of $C_2$ along pathlines gives a formula for the stretching rate $ \gamma $, and combining such a formula with the constancy of $C_1$ gives a formula for the plane vorticity $ \omega $. The result is as follows:}
\begin{equation}
\label{sol}
    \begin{aligned}
    \gamma\left(r(r_0, t), t\right) &= \frac{\gamma_0(r_0)\dot{S}(t)}{1 + \gamma_0(r_0) S(t)} - \frac{\ddot{S}(t)}{2\dot{S}(t)}, \\
    \omega\left(r(r_0, t), t\right) &= \frac{\omega_0(r_0)\left[1 + \gamma_0(r_0) S(t)\right]}{\sqrt{\dot{S}(t)}},
\end{aligned}
\end{equation}
where $ \gamma_0$ and $ \omega_0 $ denote the initial values of the stretching rate and plane vorticity, respectively: {$\gamma_0(r):= \gamma(r,0)$ and $\omega_0(r):= \omega(r,0)$},  {and we have used $\dot{S}(0)=1, \,\, S(0)=\ddot{S}(0)=0$.} These solutions explicitly describe the evolution of vorticity and its stretching rate in terms of the function $ S(t) $, which governs the temporal dynamics of the system.\\

\begin{remark}
 {Along the pathlines,} the quantities $\gamma$, $\omega$, and $z$ are all functions  {of the initial conditions} and of a single time-dependent variable $S(t)$, which is a strictly increasing function of time $t$. The evolution of $S(t)$ is governed by the following ordinary differential equation  {(to be shown in theorem \ref{path})}:
\begin{equation}
    \label{ode}
\dot{S}(t) = \left\langle \frac{1}{1 + \gamma_0(\cdot) S(t)} \right\rangle_0^{-2}, \quad S(0) = 0, \quad \dot{S}(0) = 1,
\end{equation}
where the angular brackets denote the volume-averaged integral over the initial cylindrical domain:
\begin{equation}
\label{eq:mean0}
    \langle f(\cdot,t) \rangle_0 := \frac{1}{\pi R^2} \int_0^{2\pi} \int_0^R f(r_0,t)\, r_0 \, \mathrm{d}r_0 \, \mathrm{d}\theta_0.
\end{equation}
This average is taken over the Lagrangian coordinates $(r_0, \theta_0)$ at the initial time, distinguishing it from the Eulerian average \eqref{eq:average} over current spatial coordinates.  {Despite the presence of an integral, equation \eqref{ode} is an ordinary differential equation for the scalar function $ S(t) $, as the integration is performed over the fixed initial data $ \gamma_0(r_0) $ and yields a deterministic function of $ S(t) $ alone. Since $\gamma_0$ is continuous on the compact interval $[0,R]$, the map $S \mapsto \langle (1 + \gamma_0 S)^{-1} \rangle_0$ is smooth for $S < -1/\min \gamma_0$ (where $-1/\min \gamma_0 > 0$ because $\gamma_0 \not\equiv 0$ and  $\langle\gamma_0\rangle_0 = 0$), and hence its square {multiplicative} inverse is locally Lipschitz {for $t\in [0,T_*)$, where $T_*$ is the so-called singularity time, defined to be the first time at which $S = -1/\min \gamma_0$. It obviously satisfies $t_*\leq T_*$}. The Picard–Lindel\"of theorem thus guarantees a unique local solution.} 

{Consequently, on the entire regularity interval $[0, T_*)$, we have $1+\gamma_0 S > 0$, $\dot{S}(t)$ is continuous and strictly positive, and
\begin{equation}
    A(t) = \dot{S}(t)^{-1/2} > 0 \quad \text{for all } t \in [0, T^*).
\end{equation}
Thus, $A^{-1} \in C^0([0,T_*)) \subset L^2_{\mathrm{loc}}(0,T_*)$, extending the validity of the definition of $S(t)$ in equation \eqref{eq:AtoS} from the local time $t_*$ to the singularity time $T_*$. Once $S(t)$ is determined on $[0,T_*)$, it is straightforward to verify that the Lagrangian evolution of $\gamma$, $\omega$, and the pathlines in \eqref{sol} and Theorem~\ref{path} can be fully extended to the regularity interval $[0, T_*)$. {From here on, unless otherwise specified, all time-dependent functions are defined on the whole regularity time interval $[0, T_*)$.}}

{From equation} \eqref{ode}, the second derivative of $S(t)$ is:
\begin{equation}
    \label{dds}
\ddot{S}(t) = 2 \dot{S}(t)^{\frac{5}{2}} \left\langle \frac{\gamma_0(r_0)}{(1 + \gamma_0(r_0) S(t))^2} \right\rangle_0.
\end{equation}
The derivation of equation \eqref{ode} will be presented in the proof of Theorem~\ref{path} below  {by integrating formula \eqref{ss} from $r_0=0$ to $R$ and applying the no-flow boundary condition $u_r(r=R,t)$. This equation is known in a slightly different context since the work by \cite{constantin2000euler}, where he uses the notation $\tau$ instead of $S$: see equation (19) there. In its modern form, equation \eqref{ode} was first introduced by \cite{mulungye2015symmetry} (see equation (3.4) there).}
\end{remark}

A fundamental aspect of the Lagrangian formulation of fluid dynamics is the determination of pathlines, which describe the trajectories of fluid particles over time. Based on the assumptions and velocity structure introduced earlier, we derive  explicit expressions for the pathline functions:

\begin{theorem}[Pathline Solution and Lagrangian Velocity Field]
    \label{path}
    Consider a three-dimensional incompressible Euler fluid with velocity field given by equation \eqref{velo}. Under the Lagrangian description, the radial, vertical, and angular components of the pathline are expressed as follows:
    \begin{equation}
        \label{pathline}
    \begin{aligned}
        r(r_0, t) &= \dot{S}(t)^{\frac{1}{4}} \left[ \int_0^{r_0} \frac{2 r_0'}{1 + \gamma_0(r_0') S(t)} \, \mathrm{d}r_0' \right]^{\frac{1}{2}}, \\
        z(r_0, z_0, t) &= \frac{z_0 \left[1 + \gamma_0(r_0) S(t)\right]}{\sqrt{\dot{S}(t)}}, \\
        \theta(r_0, \theta_0, t) &= \theta_0 + \left( \int_0^{r_0} \omega_0(r_0') r_0' \, \mathrm{d}r_0' \right) \cdot \left( \int_0^{t} \frac{1}{r(r_0, t')^2} \, \mathrm{d}t' \right),
    \end{aligned}
    \end{equation}
    where $r_0 = r(t=0)${, $z_0 = z(t=0)$} and $\theta_0 = \theta(t=0)$ denote the initial radial, {vertical} and angular positions of the fluid particle at $t = 0$.

Furthermore, the Lagrangian velocity field can be derived from the pathline solutions and is given by the following expressions:
\begin{equation}
    \begin{aligned}
    \label{eq:Lagr_vel}
    u_r(r(r_0, t), t) &= \frac{\int_0^{r_0} \frac{(1 + \gamma_0 S)\ddot{S} - 2\gamma_0 \dot{S}^2}{(1 + \gamma_0 S)^2} r_0' \, \mathrm{d}r_0'}{2 \dot{S}^{\frac{3}{4}} \left( \int_0^{r_0} \frac{2 r_0'}{1 + \gamma_0 S} \, \mathrm{d}r_0' \right)^{\frac{1}{2}}}, \\
    u_{\theta}(r(r_0, t), t) &= \frac{\int_0^{r_0} \omega_0(r_0') r_0' \, \mathrm{d}r_0'}{\left[ \int_0^{r_0} \frac{2 r_0' \sqrt{\dot{S}}}{1 + \gamma_0 S} \, \mathrm{d}r_0' \right]^{\frac{1}{2}}}, \\
    u_z(r(r_0, t),z(r_0,z_0,t), t) &= z_0 \gamma_0 \sqrt{\dot{S}} - \frac{1}{2} z_0 \ddot{S} \dot{S}^{-\frac{3}{2}} (1 + \gamma_0 S).
\end{aligned}
\end{equation}

\end{theorem}

\vspace{12pt}
\noindent \textit{Proof:} The expression for $z(r_0, z_0, t)$ is directly derived from the conservation law associated with the constant of motion $C_3$ in equation \eqref{constantofmotion}. {Thus,} the core of the proof lies in determining the relationship between the radial and angular components of the pathline, $r(r_0, t)$ and $\theta(r_0, \theta_0, t)$, and their initial values.

To this end, we consider the Jacobian determinant of the transformation from {$(r(r_0,t), \theta(r_0,t))$} to the initial coordinates $(r_0, \theta_0)$, defined as:
\begin{equation}
    \label{J}
J = \frac{\partial (r(r_0,t), \theta(r_0, \theta_0,t))}{\partial (r_0, \theta_0)} =
\begin{vmatrix}
\frac{\partial r}{\partial r_0} & \frac{\partial r}{\partial \theta_0} \\
\\
\frac{\partial \theta}{\partial r_0} & \frac{\partial \theta}{\partial \theta_0}
\end{vmatrix}
=
\begin{vmatrix}
\frac{\partial r}{\partial r_0} & 0 \\
\\
\frac{\partial \theta}{\partial r_0} & 1
\end{vmatrix}
= \frac{\partial r}{\partial r_0},
\end{equation}
where the relations $\frac{\partial r}{\partial \theta_0} = 0$ and $\frac{\partial \theta}{\partial \theta_0} = 1$ follow  from axisymmetry.

 {The evolution equation for the Jacobian determinant follows from its definition as the rate of change of area under the flow map. Specifically, differentiating $J=\frac{\partial r}{\partial r_0}$ with respect to time and using the radial pathline evolution equation $\mathrm{d}r/\mathrm{d}t = u_r(r,t)$ yields} 
\begin{equation}
    \label{Jaco}
\frac{\dot{J}}{J} = \frac{\partial u_r}{\partial r}, \quad J(0) = 1,
\end{equation}
which is a first-order linear ordinary differential equation.

 {To derive the radial pathline solution $r(r_0,t)$, we note that the LHS of the Jacobian evolution equation \eqref{Jaco} is a material time derivative. It turns out that the RHS of that equation is also a material time derivative. This will allow us to find a first integral. To see how the RHS is a material time derivative, we use the incompressibility condition \eqref{gammatour} to get $\frac{\partial u_r}{\partial r} = -\gamma-\frac{u_r}{r}$. Now, from the Lagrangian solution \eqref{sol} it is clear that $\gamma$ is a material time derivative. Finally, from the pathline definition $\dot{r} = u_r(r,t)$, we get $\frac{u_r}{r} = \frac{\dot{r}}{r}$, again a material time derivative. Substituting these relations, integrating with respect to time from $0$ to $t$, and then exponentiating, yields the following ordinary differential equation for the radial map:}
\begin{equation}
    \label{ss}
\frac{r}{r_0} \frac{\partial r}{\partial r_0} = \frac{\sqrt{\dot{S}(t)}}{1 + \gamma_0(r_0) S(t)}.
\end{equation}
Solving this equation yields the radial pathline function $r(r_0, t)$, consistent with the expression \eqref{pathline} stated in Theorem~\ref{path}. 
 {Furthermore, using the area-preserving identity:\begin{equation}
    \iint_{(r,\theta)\in\Omega} r\,dr\,\mathrm{d}\theta = \pi R^2 = \iint_{(r_0,\theta_0)\in\Omega} J r(r_0,t)\,dr_0\,\mathrm{d}\theta_0,
\end{equation}
and replacing $J$ from equation \eqref{J}, we derive the relation:
\begin{equation}
    \label{aj}
\left\langle \frac{r}{r_0} \frac{\partial r}{\partial r_0} \right\rangle_0 = 1,
\end{equation}
where the angular brackets are defined in equation \eqref{eq:mean0}. Combining equations  \eqref{aj} and \eqref{ss}, we obtain the ordinary differential equation governing $S(t)$, which was introduced in equation \eqref{ode}.}

 {Next, we combine the angular pathline evolution equation $u_\theta = r d \theta/d t$ with the definition \eqref{urtogamma} of $u_\theta$, the result \eqref{ss} for the Jacobian, and the Lagrangian solution \eqref{sol} for $\omega$, to obtain the angular pathline function $\theta(r_0, \theta_0, t)$ via quadrature, matching the expression \eqref{pathline} stated in Theorem~\ref{path}.}

Finally, the proof of equations \eqref{eq:Lagr_vel} follows directly from the definitions $u_r = \mathrm{d}r/\mathrm{d}t, u_\theta = r \mathrm{d}\theta/\mathrm{d}t$ and $u_z = \mathrm{d}z/\mathrm{d}t$.

This completes the derivation of all components of the pathline under the Lagrangian description. \hfill $\square$\\

\begin{remark}
The Eulerian solution can be obtained by expressing the initial position $\bm{X}_0$ as a function of the current position $\bm{X}$ and time $t$, i.e., $\bm{X}_0 = \bm{X}_0(\bm{X}, t)$, and substituting this into the Lagrangian solution. However, for the    {Gibbon-Fokas-Doering model}, the mapping $\bm{X}_0 = \bm{X}_0(\bm{X}, t)$ is implicit due to the structure of the pathline equations  \eqref{pathline}. As a result, the Eulerian solution cannot generally be expressed in an explicit closed form. Nevertheless, under certain specific Lagrangian solutions, it is possible to derive explicit Eulerian expressions, one of which will be presented in  section \ref{parabola}.
\end{remark}

\begin{remark}
From equation \eqref{ode}, we observe that the temporal evolution of the function $S(t)$ depends solely on the initial stretching rate $\gamma_0(r_0)$. Consequently, Theorem~\ref{path} implies that each pair of Lagrangian vorticity and stretching rate $(\gamma, \omega)$ uniquely determines a corresponding horizontal pathline $(r, \theta)$, a property analogous to that of the Eulerian description. For clarity, we define the pair $(\gamma, \omega)$ as the \textbf{vorticity coordinate}. Based on this observation, once the vorticity coordinate is determined, the full pathline solution for the    {Gibbon-Fokas-Doering model} becomes known. Therefore, our subsequent analysis will focus exclusively on the dynamics of the vorticity coordinate.
\end{remark}

\begin{remark}
There are two notable special cases in the pathline solution \eqref{pathline} of the    {Gibbon-Fokas-Doering model}:
\begin{itemize}
    \item[a)] \textbf{Pathlines starting at the center}: If the initial radial position is $r_0 = 0$, then the radial component of the pathline satisfies:
\begin{equation}
    r(r_0 = 0, t) = 0,
\end{equation}
indicating that such pathlines remain on the central axis ($ r = 0 $) for all time;

    \item[b)] \textbf{Pathlines starting at the boundary}: For pathlines originating from the cylindrical boundary at $r_0 = R$, we obtain:
\begin{equation}
    r(r_0 = R, t) = R, \quad \theta(r_0 = R, t) = \theta_0 + \frac{1}{2}\langle \omega_0 \rangle_0 t.
\end{equation}
   This shows that boundary pathlines do not leave the boundary and rotate with a constant angular velocity equal to $\frac{1}{2}\langle \omega_0 \rangle_0$. These properties are consistent with the no-flow boundary condition and the axisymmetric nature of the flow.
\end{itemize}

\end{remark}

\begin{remark}
According to the pathline solutions given in equations \eqref{sol} and \eqref{pathline}, the vertical component of the pathline satisfies:
\begin{equation}
    z(r_0, z_0, t) = -z(r_0, -z_0, t),
\end{equation}
which indicates that the motion in the $z$-direction is symmetric about the plane $z = 0$. In contrast, the radial and angular components $r(r_0, t)$ and $\theta(r_0, \theta_0, t)$ are independent of $z_0$. This implies that pathlines starting from the same horizontal position (i.e., the same $r_0$ and $\theta_0$) exhibit identical behavior in the $\hat{r}$ and $\hat{\theta}$ directions, regardless of their initial vertical position.\\
\end{remark}

\section{Singularity and Asymptotic Behavior of    {Gibbon-Fokas-Doering model}} \label{singular}

We observe that the stretching rate $\gamma$, as defined by equation \eqref{sol}, may potentially become infinite. This occurs when the condition
\begin{equation}\
\label{eq:singularCondition}
    1 + \gamma_0(r_0) S(t) = 0
\end{equation}
is satisfied for some $r_0$ and $t$, even if the initial data for $\gamma$ and $\omega$ are smooth. The specific values of $r_0$ and $t$ at which this condition is met are referred to as the \textbf{blowup point}, denoted $r_0 = r_-$, and the \textbf{singularity time}, denoted $t = T_*$, respectively. That is,
\begin{equation}
    1 + \gamma_0(r_-) S(T_*) = 0.
\end{equation}

 {
\begin{remark}
    In principle, the condition \eqref{eq:singularCondition} does not necessarily imply a blowup of the stretching rate $\gamma$, because there exists a special scenario in which the two terms $\displaystyle \frac{\gamma_0(r_0)\dot{S}(t)}{1 + \gamma_0(r_0) S(t)}$ and $\displaystyle \frac{\ddot{S}(t)}{2\dot{S}(t)}$ diverge at exactly the same rate, leading to their cancellation and leaving $\gamma$ regular, see the expression of $\gamma$ in \eqref{sol}. However, such a cancellation does not occur in the analysis presented below; hence, within the context of this paper, the fulfillment of \eqref{eq:singularCondition} indeed implies a genuine blowup of $\gamma$.
\end{remark}
}

The singularity time $T_*$ can be either finite or infinite. An infinite value of $T_*$ implies the absence of a finite-time singularity. 
 {At each time $t>0$, the fact $S(t)>0$ implies: $1 + \gamma_0(r_0) S(t) < 1 + \gamma_0(r_1) S(t)$ if and only if $\gamma_0(r_0)<\gamma_0(r_1)$. Now, because $S(0)=0$ the functions of time $h(r_0,t) := 1 + \gamma_0(r_0) S(t)$ all start at the value $1$ at $t=0$: $h(r_0,0) = 1$. These functions are ordered by the previous result, namely: when $t>0$ we have $h(r_0,t) < h(r_1,t)$ if and only if  $\gamma_0(r_0)<\gamma_0(r_1)$. As these functions are continuous functions of time $t$ (because $S$ is continuous), the first function that can take the value $0$ is the one for which $\gamma_0(r_0)$ is minimal, and we will have $1 + \min_{r_0 \in [0,R]}\gamma_0(r_0) S(T_*) = 0$, with $T_*$ the time at which this happens. As $S(T_*)>0$, a necessary condition for a finite-time singularity is $\min_{r_0 \in [0,R]}\gamma_0(r_0) < 0$, which is satisfied in our case (except in the trivial case $\gamma_0 \equiv 0$) because $\langle \gamma_0 \rangle_0 = 0$. 
We denote a point where $\gamma_0(r_0)$ attains its minimum value as $r_- := \arg\min_{r_0 \in [0,R]} \gamma_0(r_0)$. 
We conclude that \textbf{any blowup must occur at the pathline starting at such a point $r_-$, where $\gamma_0(r_0)$ attains its minimum value}.} This leads us to define the \textbf{blowup pathline} as the fluid trajectory starting from the minimum point $r_0 = r_-$. 

To further analyze the existence and magnitude of a finite singularity time $T_*$, we examine the relationship between $S(t)$ and $\gamma_0(r_0)$. From the blowup condition above, we deduce the corresponding critical value of $S$ at the singularity:
\begin{equation}
    S_* := S(T_*) = -\frac{1}{\inf_{r_0 \in [0,R]} \gamma_0(r_0)},
\end{equation}
where $\inf_{r_0 \in [0,R]} \gamma_0(r_0)$ denotes the global minimum of the initial stretching rate over the domain $[0, R]$.

 {Integrating the evolution equation \eqref{ode} for $S(t)$ by separation of variables yields an explicit expression for the singularity time:}
\begin{equation}
    \label{t*}
T_* = \int_0^{S_*} \frac{1}{\dot{S}} \, \mathrm{d}S = \int_0^{S_*} \left\langle \frac{1}{1 + \gamma_0(r_0) S} \right\rangle_0^2 \mathrm{d}S,
\end{equation}
where $\langle \cdot \rangle_0$ denotes the volume average over the initial cylindrical domain is defined by equation \eqref{eq:mean0}.
From the ordinary differential equation governing $S(t)$, it follows that the function $S(t)$ is completely determined by the initial profile of $\gamma_0(r_0)$. A direct inspection of equation \eqref{t*} confirms that the nature of the singularity—whether it forms in finite time or not—is also dictated solely by the properties of $\gamma_0(r_0)$.

In particular, the singularity time satisfies the following scaling relation:
\begin{equation}
    \label{meanT}
T_*(k\gamma_0) = \frac{1}{k} T_*(\gamma_0), \quad k > 0,
\end{equation}
which shows that proportionally scaling the initial stretching rate $\gamma_0(r_0)$ results in an inverse proportional change in the singularity time $T_*$. Thus, a larger (in absolute value) negative minimum of $\gamma_0(r_0)$ leads to an earlier blowup.

It is important to emphasize that our analysis assumes regular initial conditions. We begin by examining the case where $\gamma_0(r_0)$ takes the form of an exact parabola, and then proceed to more general cases where $\gamma_0(r_0)$ is continuous but not necessarily differentiable near the blowup point.

\subsection{Parabolic Initial Stretching Rate $\gamma_0(r_0)$} \label{parabola}

In this subsection, we consider a specific example where the initial stretching rate $\gamma_0(r_0)$ takes the form of an exact parabola satisfying the area-averaged condition $\langle \gamma_0 \rangle = 0$:
\begin{equation}
    \label{exg}
\gamma_0(r_0) = {\gminus} \left(1 - \frac{2}{R^2} r_0^2 \right),
\end{equation}
where ${\gminus} \neq 0$ is a constant. In this case, the minimum of $\gamma_0$ occurs at either $r_0 = 0$ or $r_0 = R$, depending on the sign of ${\gminus}$. This choice offers several advantages:

\begin{itemize}
    \item[a)] The profile of $\gamma_0$ is controlled by a single parameter ${\gminus}$, and its minimum is clearly defined;
    
    \item[b)] The minimum and supremum of $\gamma_0$ are negatives of each other, leading to the same value for the minimum: $\min \gamma_0 = -|{\gminus}|$, regardless of whether ${\gminus}$ is positive or negative;

    \item[c)] The integrals involved in the Lagrangian solutions can be evaluated explicitly, simplifying the analysis.
\end{itemize}

Firstly, substituting this form of $\gamma_0(r_0)$ into the 
 {expressions for $\dot{S}(t)$ and $\ddot{S}(t)$ given in equations \eqref{ode} and \eqref{dds}, we obtain:}
\begin{align}
\dot{S}(t) &= 4 {\gminus}^2 S(t)^2 \left[ \ln \left| \frac{1 - {\gminus} S(t)}{1 + {\gminus} S(t)} \right| \right]^{-2}\label{rs}\,,\\[2pt]
\ddot{S}(t) &=  {32\gminus^4S(t)^3\left[ \ln \left| \frac{1 - {\gminus} S(t)}{1 + {\gminus} S(t)} \right| \right]^{-5}\left[\ln \left| \frac{1 - {\gminus} S(t)}{1 + {\gminus} S(t)} \right| + \frac{2\gminus S(t)}{1-\gminus^2S(t)^2}\right]}\,.\label{rss}
\end{align}
It is important to note that changing the sign of ${\gminus}$ does not alter the solution of $S(t)$, as the right-hand side of the above equation remains invariant under such a transformation. Furthermore,  {substituting formula \eqref{rs} into the singularity time formula \eqref{t*} and evaluating the resulting integral yields an explicit expression for the blowup time: }
\begin{equation}
\label{eq:Tstar}
    T_* = \frac{\pi^2}{6 |{\gminus}|},
\end{equation}
which confirms that the blowup occurs in finite time and depends inversely on the magnitude of ${\gminus}$.

Next, we compute the corresponding Lagrangian solutions for the vorticity coordinate $(\gamma, \omega)$. Substituting the parabolic $\gamma_0(r_0)$ from equation \eqref{exg} into the general expressions for the Lagrangian vorticity and stretching rate given in equation \eqref{sol}, we obtain:
\begin{equation}
    \begin{aligned}
    \gamma(r(r_0,t), t) &= \frac{\dot{S} R^2}{2 {\gminus} S^2 r_0^2 - R^2 S (1 + {\gminus} S)} + \frac{\dot{S}}{S} - \frac{\ddot{S}}{2 \dot{S}}, \\
    \omega(r(r_0,t), t) &= \omega_0(r_0) \left( -\frac{2 {\gminus} S}{\sqrt{\dot{S}} R^2} r_0^2 + \frac{1 + {\gminus} S}{\sqrt{\dot{S}}} \right).
\end{aligned}
\end{equation}
The radial pathline function $r(r_0, t)$ and its inverse $r_0(r, t)$ are derived as:
\begin{equation}
\label{eq:r_plot}
\begin{aligned}
    r(r_0, t) &= \frac{\dot{S}^{1/4} R}{\sqrt{2 S}} \left[ -\frac{1}{{\gminus}} \ln \left| 1 - \frac{2 {\gminus} S r_0^2}{(1 + {\gminus} S) R^2} \right| \right]^{1/2}, \quad
    r_0(r, t) = \left[ \frac{R^2 (1 + {\gminus} S)}{2 {\gminus} S} \left( 1 - \mathrm{e}^{- \frac{2 {\gminus} S r^2}{\sqrt{\dot{S}} R^2}} \right) \right]^{1/2}.
\end{aligned}
\end{equation}
The angular and vertical components of the pathline are expressed as:
\begin{equation}
\label{eq:theta_z_plot}
\begin{aligned}
    \theta(\theta_0, r_0, t) &= \theta_0 - \frac{2 {\gminus}}{R^2} \int_0^{r_0} \omega_0(r_0') r_0' \, \mathrm{d}r_0' \cdot \int_0^t \frac{S(t')}{\sqrt{\dot{S}(t')} \ln \left| 1 - \frac{2 {\gminus} S(t')}{R^2 [1 + {\gminus} S(t')]} r_0^2 \right|} \mathrm{d}t', \\
  z(r_0, z_0, t) &= {\frac{z_0}{\sqrt{\dot{S}}} \left[1 + {\gminus} S\left(1-\frac{2 }{R^2} r_0^2\right)\right].}
\end{aligned}
\end{equation}

\begin{figure}
	\centering
	\begin{subfigure}[c]{0.46\textwidth}
		\centering
        \caption{$\gamma_0 = 10-20r_0^2,\: \omega_0 = 88$}
		\includegraphics[width=1\textwidth]{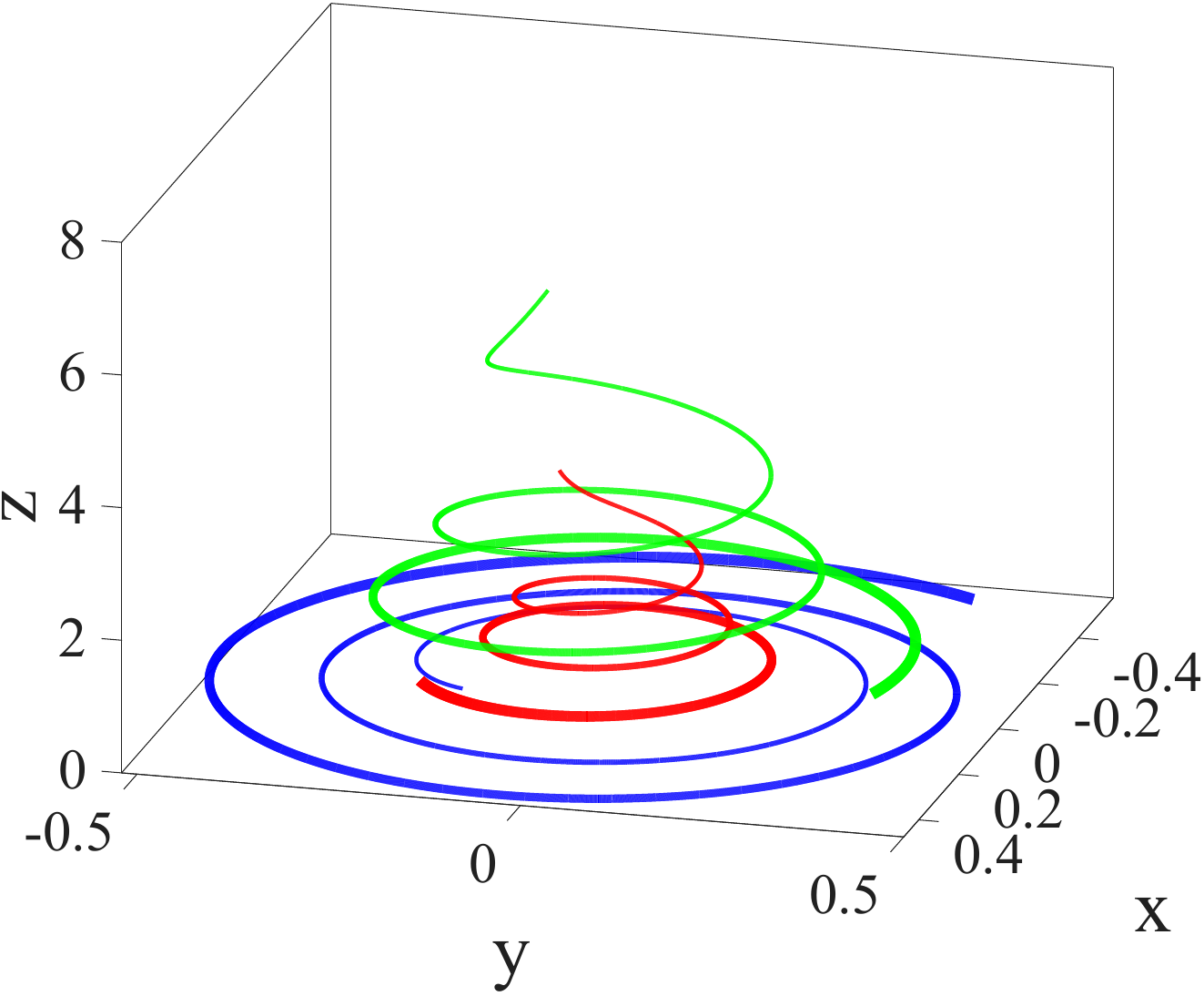}
	\end{subfigure}
        \hspace{0.05\textwidth} 
	\begin{subfigure}[c]{0.46\textwidth}
		\centering
        \caption{$\gamma_0 = 20r_0^2 - 10,\: \omega_0 = 380$}
		\includegraphics[width=1\textwidth]{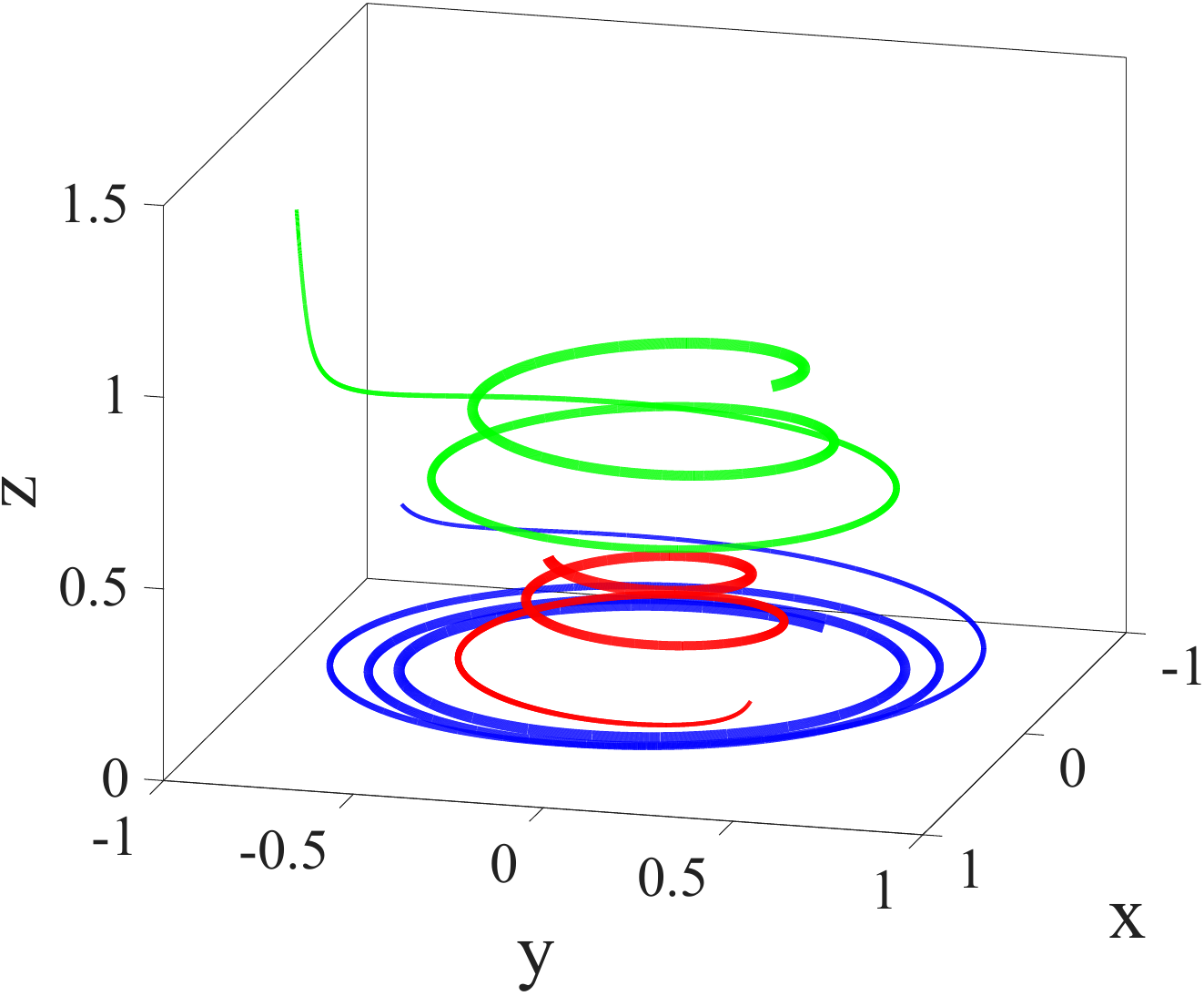}
	\end{subfigure}
    \captionsetup{justification = raggedright, singlelinecheck = false}
    \caption{Lagrangian pathlines for the Gibbon–Fokas–Doering model with parabolic initial stretching rate $\gamma_0(r_0) = f\left(1 - \frac{2r_0^2}{R^2}\right)$ and uniform initial plane vorticity $\omega_0$, for cylinder radius $R = 1$. In both panels (a) and (b), trajectories are shown for three initial positions: $(r_0, \theta_0, z_0) = (0.25, 1.64\pi, 0.4)$ [red], $(0.4, 0.38\pi, 0.9)$ [green], $(0.6, 0.81\pi, 0.1)$ [blue]. The width of each trajectory is artificially made proportional to the time to the singularity, $T_* - t$, where $T_* = \pi^2/6|f|$. (a) $f = 10$ (blowup at boundary $r_- = R$) and $\omega_0 = 88$; all trajectories have a helical ascent (theoretically reaching the cylinder axis $r=0$ at an unbounded height $z\to\infty$ at the singularity time $t=T_*$). (b) $f = -10$ (blowup at center $r_- = 0$) and $\omega_0 = 380$; all trajectories have an initial helical descent followed by a helical ascent (theoretically reaching the cylinder boundary $r=R$ at an unbounded height $z\to\infty$ at the singularity time $t=T_*$); because the red curve starts closer to the center than the other curves, it takes longer to reverse its direction from descent to ascent.}\label{fig:pathline}
\end{figure}

{Following the explicit Lagrangian solutions derived above, Figure \ref{fig:pathline} provides a concrete visualization of the Gibbon–Fokas–Doering model’s pathline dynamics for the parabolic initial stretching rate, with enhanced values of the initial vorticity (chosen uniform for simplicity) to better visualize the particles' rotation: $\omega_0 = 88$ in the case $f=10$ (singularity at the boundary, left panel) and $\omega_0 = 380$ in the case $f=-10$ (singularity at the center, right panel). The plots illustrate how fluid particles undergo spiral motion, offering an intuitive grasp of the flow geometry dictated by Theorem \ref{path}. Notably, the right panel ($f=-10$; blowup is at the center) reveals a non-monotonic vertical motion for the trajectories that start close enough to the center (red and green trajectories in the figure): they initially descend before reversing direction and ascending. This behavior arises directly from the competition between the two terms in the vertical velocity expression \eqref{eq:Lagr_vel}, and can also be understood by looking at the $z$-component of equation \eqref{eq:theta_z_plot}, which shows that for $r_0<R/\sqrt{2}$ the particle's height can have a non-monotonic behavior.}

Furthermore, the Eulerian solutions for $\gamma(r, t)$ and $\omega(r, t)$ can also be written explicitly. Substituting the parabolic $\gamma_0(r_0)$ into the vorticity coordinate equations, we obtain:
\begin{equation}
\label{eq:parabolaGamma}
    \begin{aligned}
    \gamma(r, t) &= -\frac{\dot{S}}{S (1 + {\gminus} S)} \exp\left( \frac{2 {\gminus} S r^2}{\sqrt{\dot{S}} R^2} \right) + \frac{\dot{S}}{S} - \frac{\ddot{S}}{2 \dot{S}}, \\
    \omega(r, t) &= \omega_0(r_0(r,t)) \frac{1 + {\gminus} S}{\sqrt{\dot{S}}} \exp\left( -\frac{2 {\gminus} S r^2}{\sqrt{\dot{S}} R^2} \right).
\end{aligned}
\end{equation}
Using these expressions and substituting into equation \eqref{urtogamma}, we derive the velocity components in the Eulerian frame:
\begin{equation}
    \begin{aligned}
    u_r(r, t) &= \frac{\dot{S}^{3/2} R^2}{4 {\gminus} S^2 (1 + {\gminus} S)} \cdot \frac{1}{r} \left( 1 - \mathrm{e}^{- \frac{2 {\gminus} S r^2}{\sqrt{\dot{S}} R^2}} \right) + \frac{1}{2} \left( \frac{\ddot{S}}{2 \dot{S}} - \frac{\dot{S}}{S} \right) r, \\
    u_\theta(r, t) &= \frac{1 + {\gminus} S}{\sqrt{\dot{S}}} \cdot \frac{1}{r} \int_0^r \omega_0(r_0(r', t)) r' \exp\left( -\frac{2 {\gminus} S r'^2}{\sqrt{\dot{S}} R^2} \right) \mathrm{d}r', \\
    u_z(r, z, t) &= z \left[ -\frac{\dot{S}}{S (1 + {\gminus} S)} \exp\left( \frac{2 {\gminus} S r^2}{\sqrt{\dot{S}} R^2} \right) + \frac{\dot{S}}{S} - \frac{\ddot{S}}{2 \dot{S}} \right].
\end{aligned}
\end{equation}
To further investigate the dynamical properties of this configuration, we now proceed to analyze the asymptotic behavior of the blowup pathline as $t \to T_*$. This will allow us to examine how the solution evolves near the singularity and determine the nature of the finite-time blowup. We will show that the blowup dynamics are governed by the local behavior of the initial stretching rate near its minimum point, and that the singularity manifests in a self-similar fashion in both the Lagrangian and Eulerian descriptions.

\begin{theorem}
\label{qua}
    If the initial stretching rate $\gamma_0$ takes the form given in equation \eqref{exg}, with a nonzero constant ${\gminus}$, then the blowup pathline exhibits the following asymptotic behavior in the Lagrangian framework:
\begin{eqnarray*}
\gamma(r_0 = r_-, t) \sim -\frac{1}{T_* - t}, \qquad
\omega(r_0 = r_-, t) &\sim& \omega_0(r_-)  {\frac{|\gminus| (T_*-t)}{\left|W_{-1}\left(-\sqrt{|\gminus|(T_* - t)}\right)\right|},}\\
S(t) &\sim& S_* - \mathrm{e}^{2 W_{-1}\left(-\sqrt{|f|(T_* - t)}\right)},
\end{eqnarray*}
where $W_{-1}$ denotes the negative branch of the Lambert $W$ function, defined by the identity $W_{-1}(z) \mathrm{e}^{W_{-1}(z)} = z$, {with $-\mathrm{e}^{-1} \leq z < 0$ and $W_{-1}(z) \leq -1$.}
\end{theorem}

\noindent \textit{Proof:} As $t \to T_*$, along the blowup pathline we have:
\begin{equation}
    1 + \gamma_0 S = 1 - |{\gminus}| S = 1 - \frac{S}{S_*},
\end{equation}
namely $S_* = 1/|{\gminus}|$.  
 {Substituting this relation into  equations \eqref{rs} and \eqref{rss}}, we derive the leading-order asymptotic expressions for the first and second time derivatives of $S(t)$:
\begin{equation}
    \label{interme}
\dot{S}(t) \sim 4 \left[ \ln\left(1 - \frac{S}{S_*}\right) \right]^{-2}, \quad
\ddot{S}(t) \sim \frac{32 \left[ \ln\left(1 - \frac{S}{S_*}\right) \right]^{-5}}{S_* - S}.
\end{equation}
 {An explicit asymptotic solution for $S(t)$ at times $t$ near the singularity time $T_*$ can be obtained via first replacing $\dot{S}$ into the formula for the singularity time:
\begin{equation}
    \begin{aligned}
        T_* - t &= \int_S^{S_*} \frac{1}{\dot{S}} \mathrm{d}S \sim \frac{1}{4}\int_S^{S_*}  \left[ \ln\left(1 - \frac{S}{S_*}\right) \right]^{2} \mathrm{d}S \\
        &\sim {\frac{1}{4}S_*\left(1 - \frac{S}{S_*}\right)\left[\ln\left(1 - \frac{S}{S_*}\right)\right]^2} = {S_*\left(\sqrt{1 - \frac{S}{S_*}}\ln\sqrt{1 - \frac{S}{S_*}}\right)^2\,,}
    \end{aligned}
\end{equation}
and then applying the definition of the negative branch of the Lambert $W$ function as in the statement of this theorem, 
yields the leading-order approximation
\begin{equation}
    \label{finals}
S(t) \sim S_* - S_* \mathrm{e}^{2 W_{-1}\left(-\sqrt{|{\gminus}|(T_* - t)}\right)}.
\end{equation}
}
Substituting  {the asymptotics \eqref{interme} and \eqref{finals}} into the Lagrangian expressions for $\gamma$ and $\omega$ given in equation \eqref{sol}, we compute the asymptotic behavior of the stretching rate along the blowup pathline:
\begin{equation}
    \gamma(r_0 = r_-, t) = \frac{\gamma_0 \dot{S}}{1 + \gamma_0 S} - \frac{\ddot{S}}{2 \dot{S}}
\sim -\frac{4 \left[\ln\left(1 - \frac{S}{S_*}\right) \right]^{-2}}{S_* - S} - \frac{4 \left[\ln\left(1 - \frac{S}{S_*}\right) \right]^{-3}}{S_* - S}
\sim -\frac{1}{T_* - t}\,,
\end{equation}
 {and} for the plane vorticity we obtain:
\begin{equation}
    \omega(r_0 = r_-, t) \sim \frac{\omega_0(r_-)(1 + \gamma_0 S)}{\sqrt{\dot{S}}}
\sim \frac{\omega_0(r_-)\left(1 - \frac{S}{S_*}\right)}{2 \left|\ln\left(1 - \frac{S}{S_*}\right)\right|^{-1}}
\sim \omega_0(r_-)  {\frac{|\gminus| (T_*-t)}{\left|W_{-1}\left(-\sqrt{|\gminus|(T_* - t)}\right)\right|}.}
\end{equation}
The proof is complete. \hfill $\square$

\begin{remark}
As noted by Gibbon and Ohkitani, a necessary condition for the breakdown of the Gibbon-type model—similar in spirit to the Beale-Kato-Majda theorem—is that the time integral of the $L^\infty$-norm of the stretching rate diverges:
\begin{equation}
    \int_0^{T_*} \left\|\gamma(r(r_0,t), t)\right\|_{L^{\infty}(\Omega_C)} \mathrm{d}t \to +\infty.
\end{equation}
This criterion is exemplified by Theorem~\ref{qua}, which demonstrates that such a  divergence occurs precisely at the blowup time $T_*$ \citep{beale1984remarks, ohkitani2000numerical}.
\end{remark}

Starting from the asymptotic behavior of $S(t)$, $\dot{S}(t)$, and $\ddot{S}(t)$ given in equations \eqref{interme} and \eqref{finals}, we now analyze the Eulerian vorticity coordinate as $t \to T_*$.

\begin{figure}
	\centering
	\begin{subfigure}[t]{0.325\textwidth}
		\centering
        \caption{$t = 0$}
		\includegraphics[width=1\textwidth]{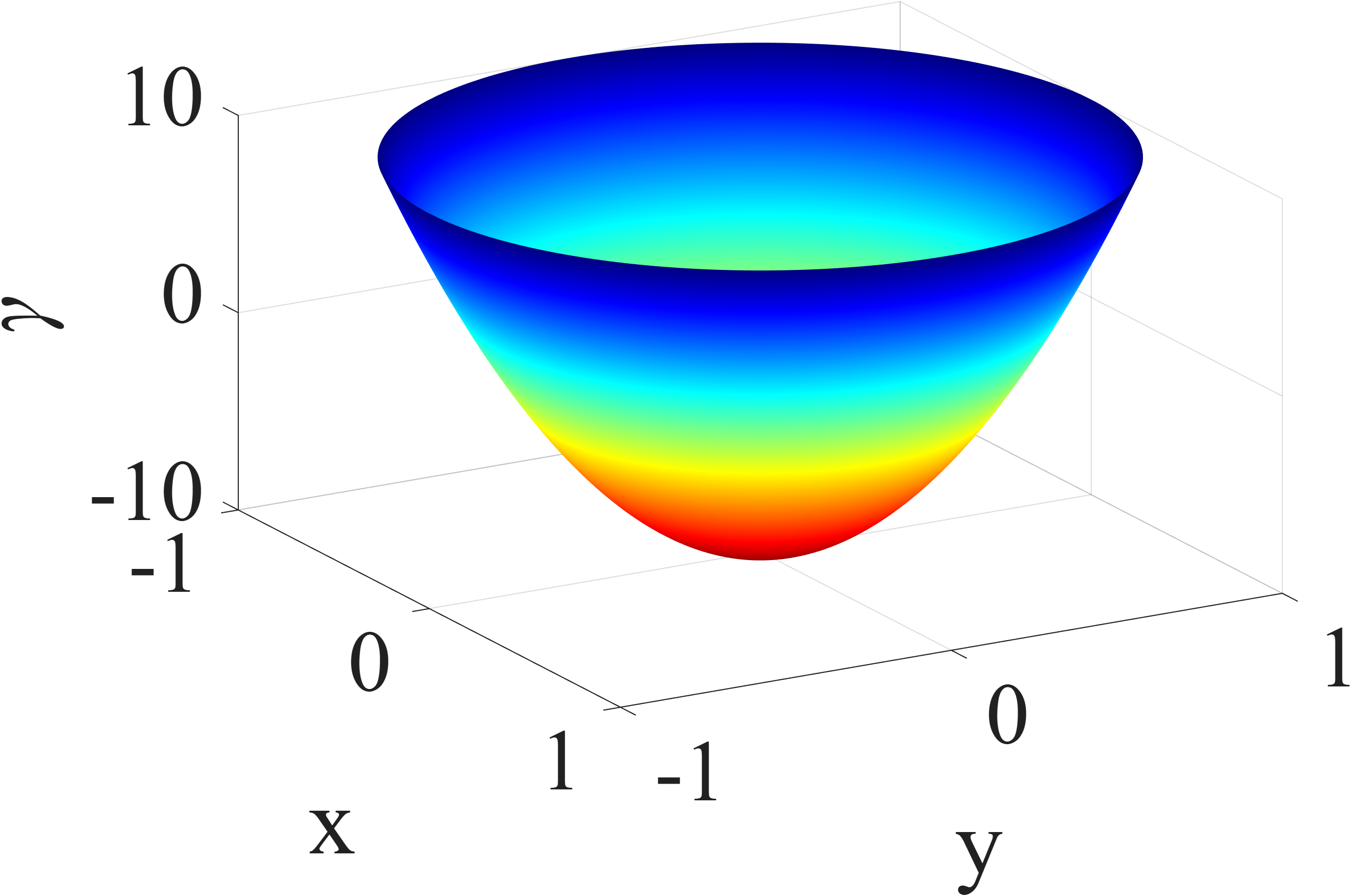}
	\end{subfigure}
	\begin{subfigure}[t]{0.325\textwidth}
		\centering
        \caption{$t = 0.0822$}
		\includegraphics[width=1\textwidth]{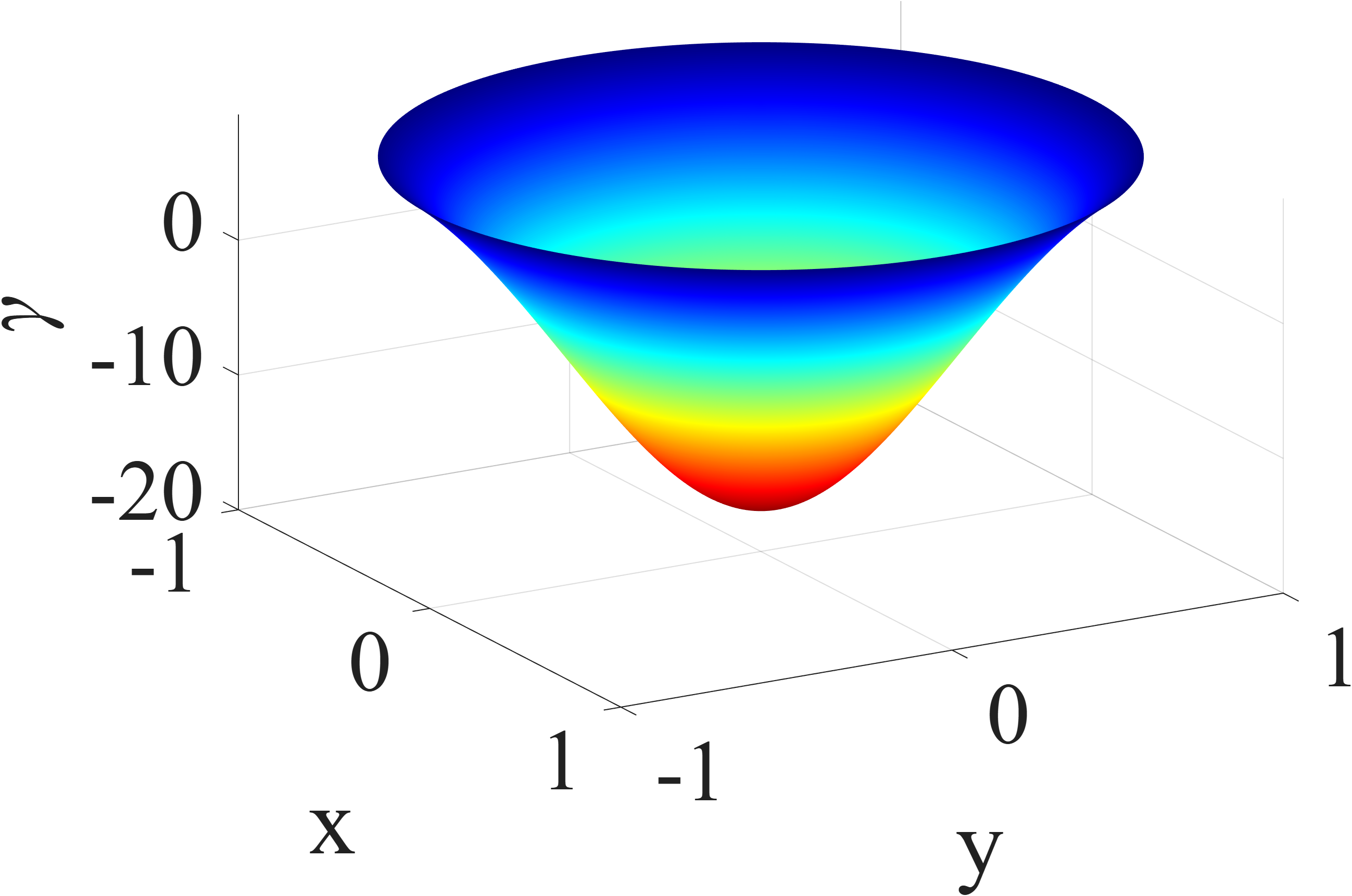}
	\end{subfigure}
	\begin{subfigure}[t]{0.325\textwidth}
		\centering
        \caption{$t = 0.1644$}
		\includegraphics[width=1\textwidth]{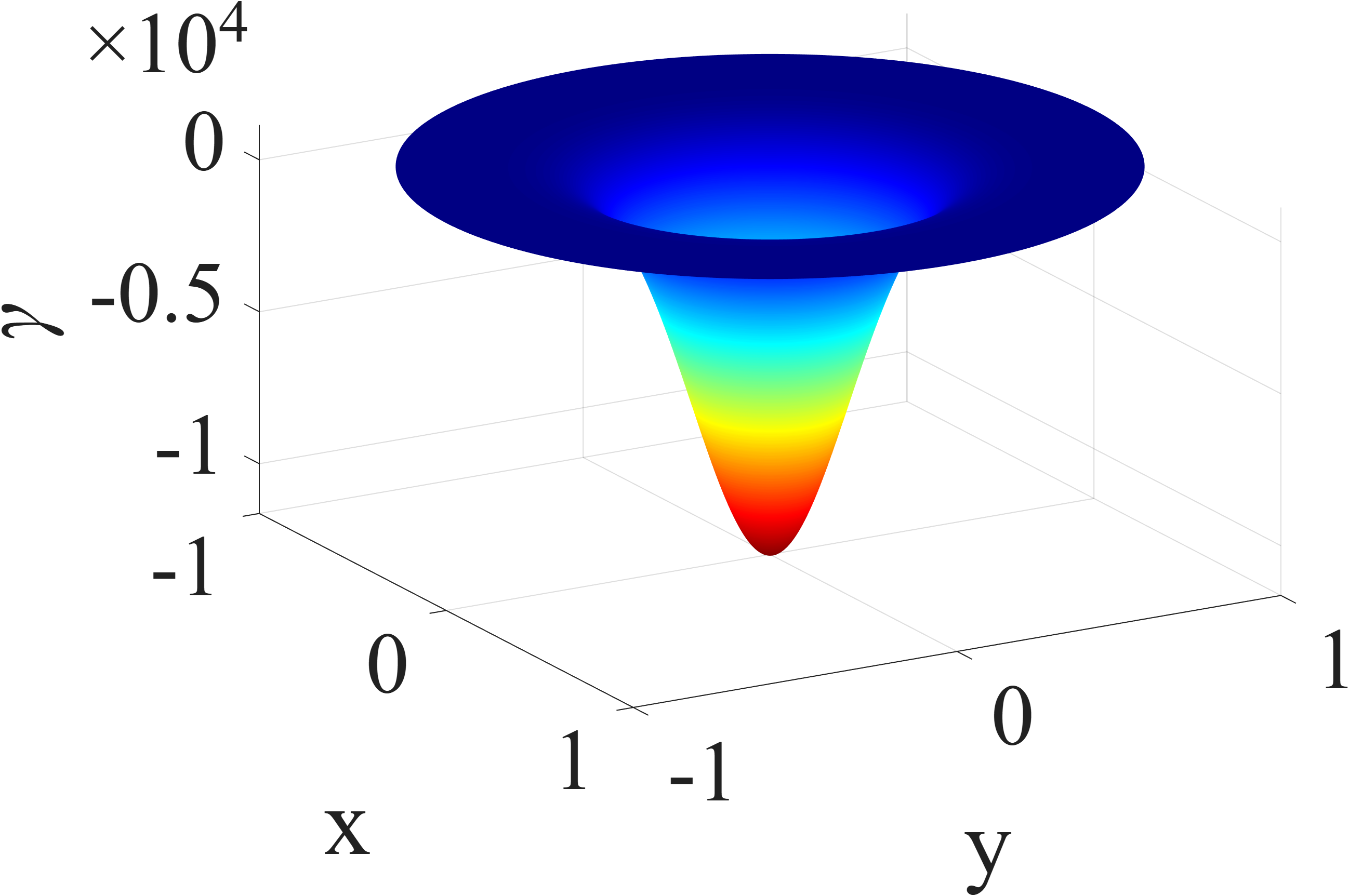}
	\end{subfigure}
        \begin{subfigure}[t]{0.325\textwidth}
		\centering
            \caption{$t = 0$}
		\includegraphics[width=1\textwidth]{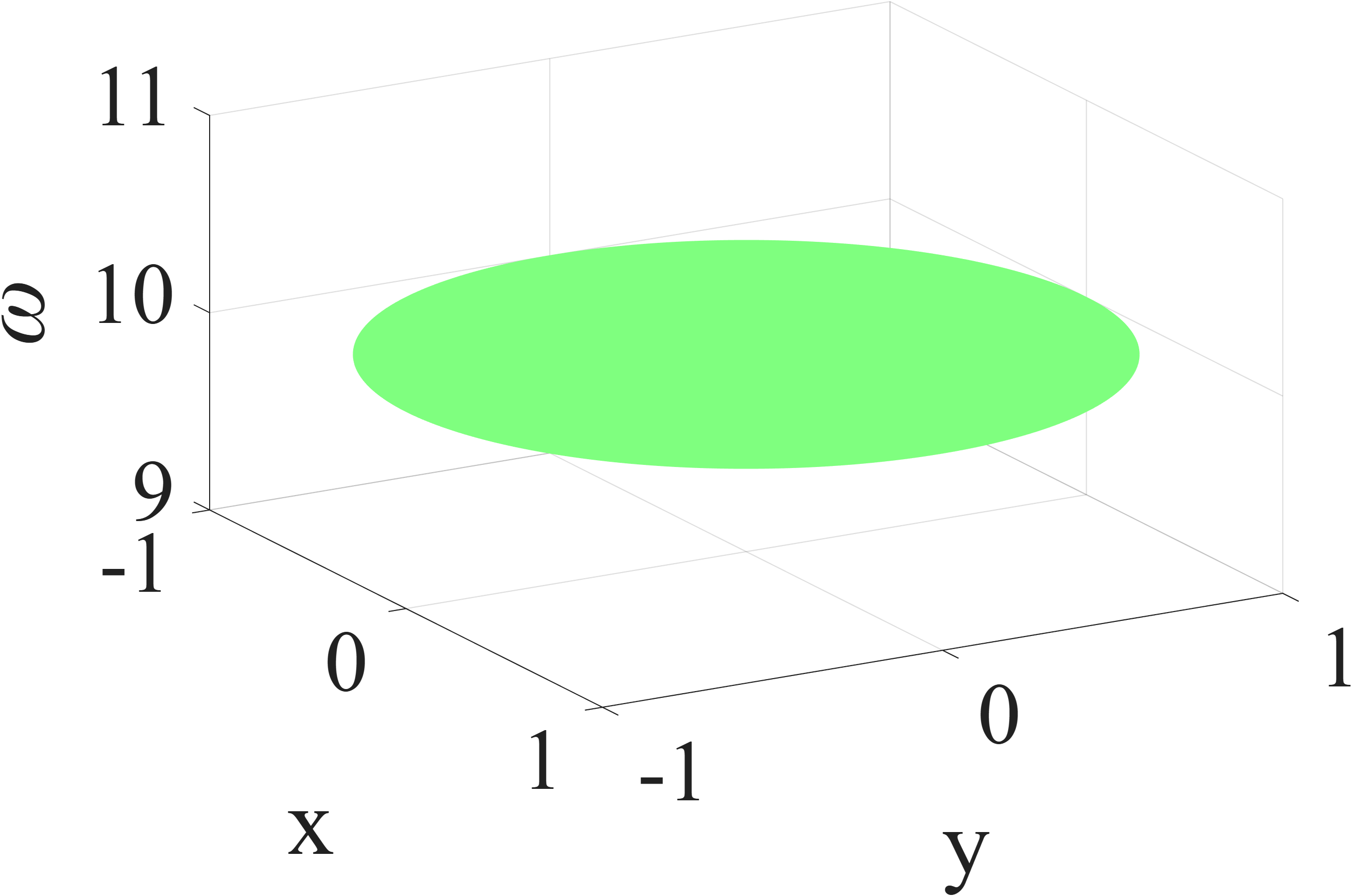}
	\end{subfigure}
        \begin{subfigure}[t]{0.325\textwidth}
		\centering
            \caption{$t = 0.0822$}
		\includegraphics[width=1\textwidth]{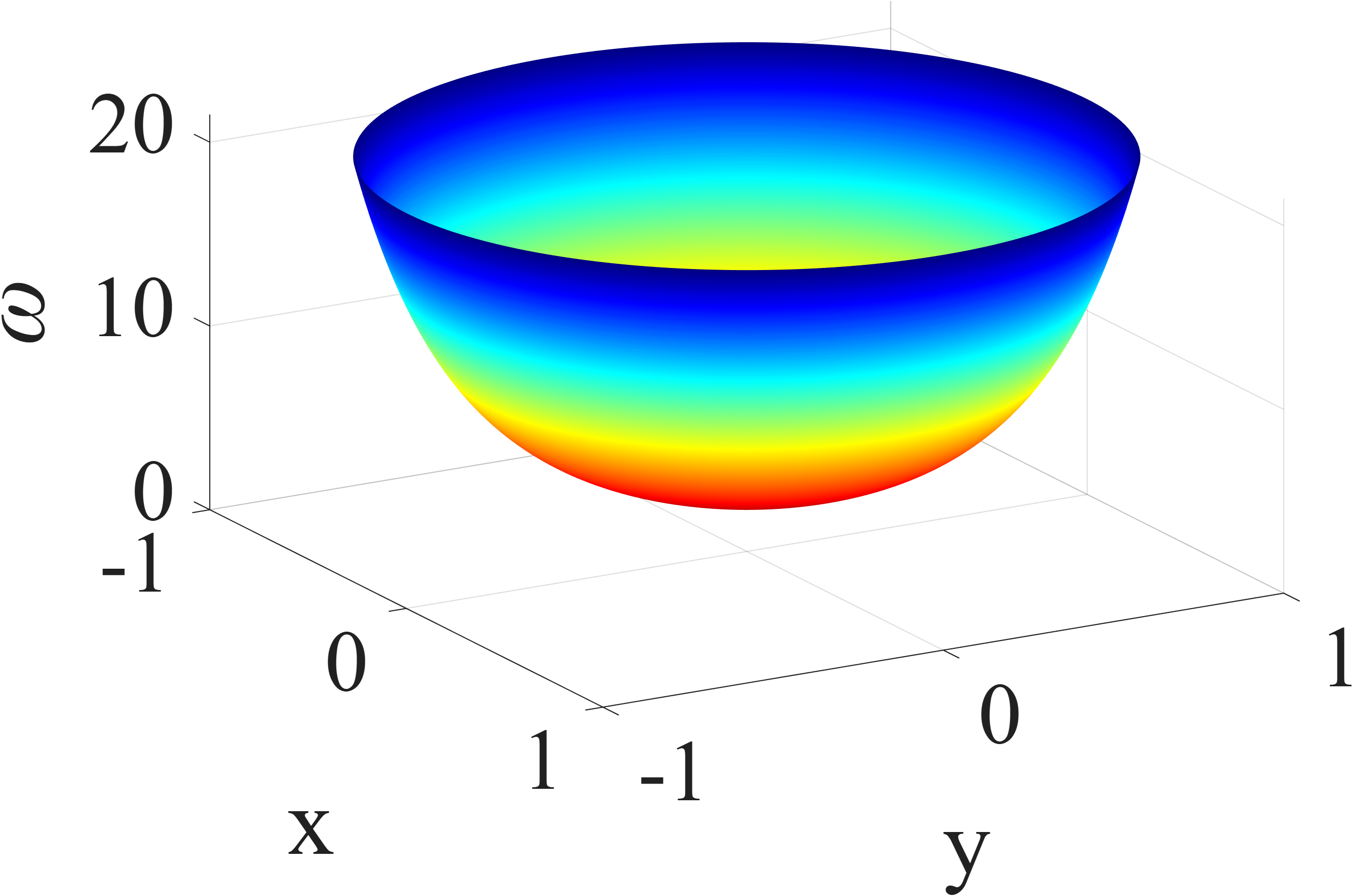}
	\end{subfigure}
        \begin{subfigure}[t]{0.325\textwidth}
		\centering
            \caption{$t = 0.1644$}
		\includegraphics[width=1\textwidth]{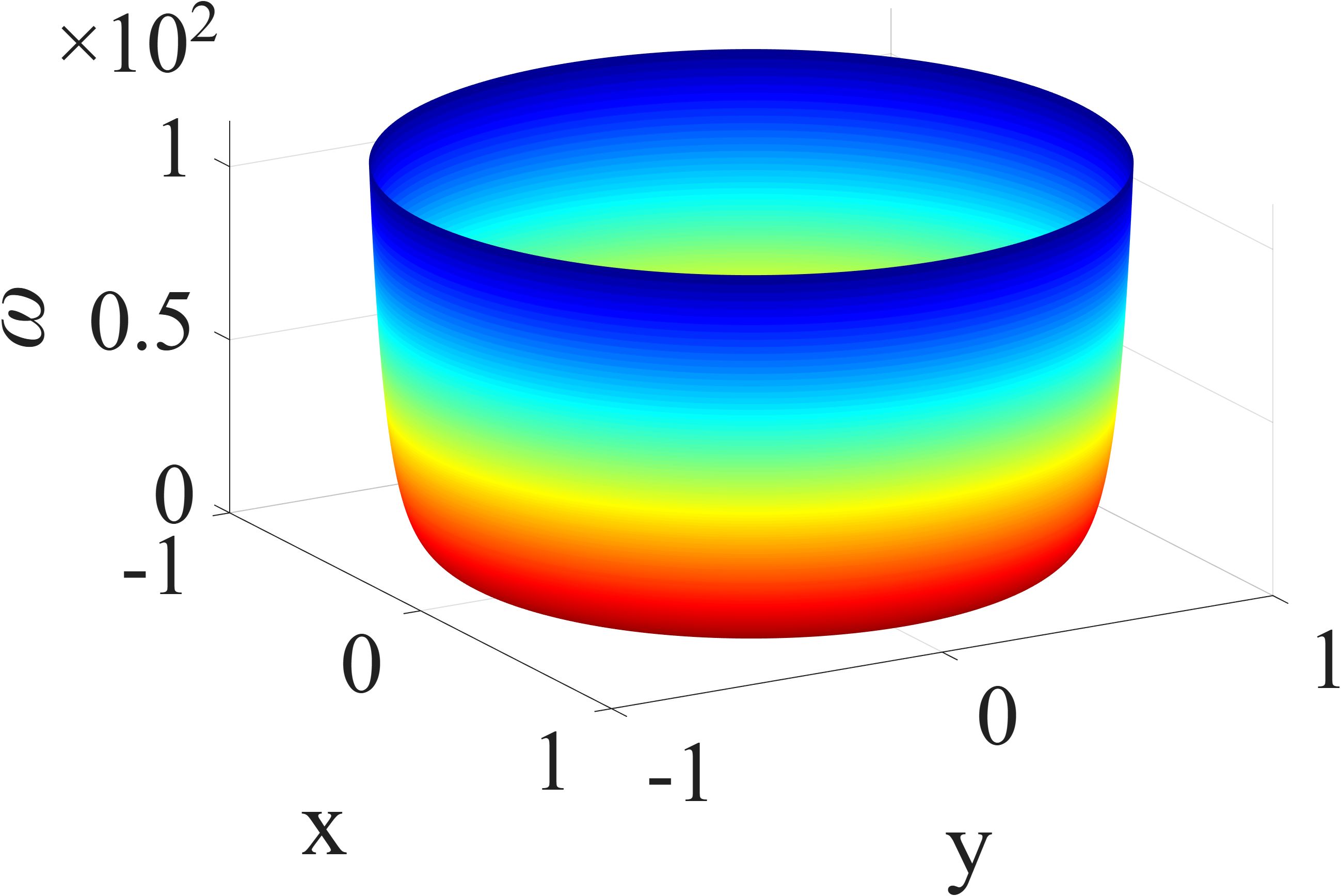}
	\end{subfigure}
    \captionsetup{justification = raggedright, singlelinecheck = false}
    \caption{Evolution of the parabolic  {initial} stretching rate $\gamma_0(r_0) = 20r_0^2 - 10$ (hence $f=-10,R=1$; {blowup is at the center})  {and the uniform initial plane vorticity $\omega_0(r_0) = 10$} in the Eulerian description. The singularity time is computed as $T_* = \frac{\pi^2}{6|{\gminus}|} \approx 0.16449$. According to equation \eqref{eq:gamma<0}, the asymptotic behavior of the stretching rate $\gamma$ is expected to take the form of a bell-shaped function, which is consistent with the spatial profile observed in this figure.  {While, at the center ($r=0$), $\gamma$ develops a finite-time singularity and $\omega$ develops a zero, at the boundary ($r=R$) $\omega$ exhibits a divergent peak, confirming their mutually exclusive blowup locations. The slower growth of $\omega$ compared to $\gamma$ (visible in the $z$-axis scale) aligns with its Lambert-W-type asymptotics derived in \eqref{eq:gamma<0}.}}\label{fig:gamma-<0}
\end{figure}

\begin{itemize}
    \item[a)] \textbf{Case ${\gminus} < 0$: Blowup at the center $r = 0$}: When ${\gminus} < 0$, the minimum point of the initial stretching rate is located at $r_- = 0$. In this case, we have:
\begin{equation}
    \lim_{t \to T_*} (1 + {\gminus} S(t)) = 0.
\end{equation}
Using the leading-order asymptotics derived in equations \eqref{interme} and \eqref{finals}, we obtain the following asymptotic forms for the Eulerian vorticity coordinate:
\begin{equation}
    \label{eq:gamma<0}
\begin{aligned}
    \gamma(r,t) &\sim -\frac{1}{T_* - t} \mathrm{e}^{-\frac{2r^2}{R^2} \left|W_{-1}\left(-\sqrt{|\gminus|(T_* - t)}\right)\right|} - \frac{1}{2(T_* - t) \left|W_{-1}\left(-\sqrt{|\gminus|(T_* - t)}\right)\right|}, \\
  \omega(r,t) &\sim \omega_0(r_0)|\gminus|^{1 - \frac{r^2}{R^2}} \left|W_{-1}\left(-\sqrt{|\gminus|(T_* - t)}\right)\right|^{\frac{2r^2}{R^2} - 1} (T_* - t)^{1 - \frac{r^2}{R^2}}.\\
\end{aligned}
\end{equation}
From the expression for $\gamma(r,t)$, we observe that the stretching rate blows up over the entire spatial domain $\Omega_C$, with the most rapid growth occurring at the origin $r = 0$, {where
$\gamma(0, t) \sim -\frac{1}{T_* - t}$}, confirming that the singularity first emerges at the center.

In contrast, the plane vorticity $\omega(r,t)$ behaves differently: it decays to zero throughout the interior of the domain but diverges at the boundary $r = R$. This boundary blowup occurs more slowly than {the blowup of stretching rate at the center}, with a scaling that is proportional to the negative branch of the Lambert $W$ function: $W_{-1}\left(-\sqrt{|{\gminus}|(T_* - t)}\right)$. Despite the slower growth rate, the presence of a finite-time singularity at the boundary confirms the non-regularity of the solution;

 {An example of the above asymptotic behavior for $\gminus<0$  and its evolution is illustrated in Figure~\ref{fig:gamma-<0} in the case ${\gminus}=-10, R=1$.}

    \item[b)] \textbf{Case ${\gminus} > 0$: Blowup at the boundary $r = R$}: For ${\gminus} > 0$, the blowup point is located at the cylindrical boundary $r_- = R$, where:
\begin{equation}
    \lim_{t \to T_*} (1 - {\gminus} S(t)) = 0, \quad \text{and} \quad \lim_{t \to T_*} (1 + {\gminus} S(t)) = 2.
\end{equation}
The corresponding asymptotic behavior of the Eulerian stretching rate and plane vorticity is given by:
\begin{equation}
    \label{eq:gamma>0}
\begin{aligned}
    \gamma(r,t) &\sim -\frac{1}{2} \gminus^{1-\frac{r^2}{R^2}} \left|W_{-1}\left(-\sqrt{|\gminus|(T_* - t)}\right)\right|^{2\left(\frac{r^2}{R^2} - 1\right)} (T_* - t)^{-\frac{r^2}{R^2}} - \frac{1}{2(T_* - t) W_{-1}\left(-\sqrt{|\gminus|(T_* - t)}\right)},
    \\
    \omega(r,t) &\sim 2\gminus^{\frac{r^2}{R^2}}\omega_0(r,t) \left|W_{-1}\left(-\sqrt{|\gminus|(T_* - t)}\right)\right|^{1 - \frac{2r^2}{R^2}} (T_* - t)^{\frac{r^2}{R^2}}.
\end{aligned}
\end{equation}
Similar to the previous case, here the stretching rate $\gamma(r,t)$ becomes singular everywhere on the spatial domain $\Omega_C = [0, R] \times [0, 2\pi) \times \mathbb{R}$ {at $t = T_*$}, but with the fastest growth rate occurring at the boundary $r = R$, where $\gamma(R, t) \sim -\frac{1}{2(T_* - t)}$.

As for the plane vorticity $\omega(r,t)$, it decays to zero at the singularity time everywhere except at the center of the domain $r=0$, where it blows up. This indicates that the singularity formation mechanism differs depending on the sign of the initial stretching rate ${\gminus}$, even though both cases share the same minimum value and singularity time.
\end{itemize}

\begin{figure}
	\centering
	\begin{subfigure}[t]{0.325\textwidth}
		\centering
            \caption{$t = 0$}
		\includegraphics[width=1\textwidth]{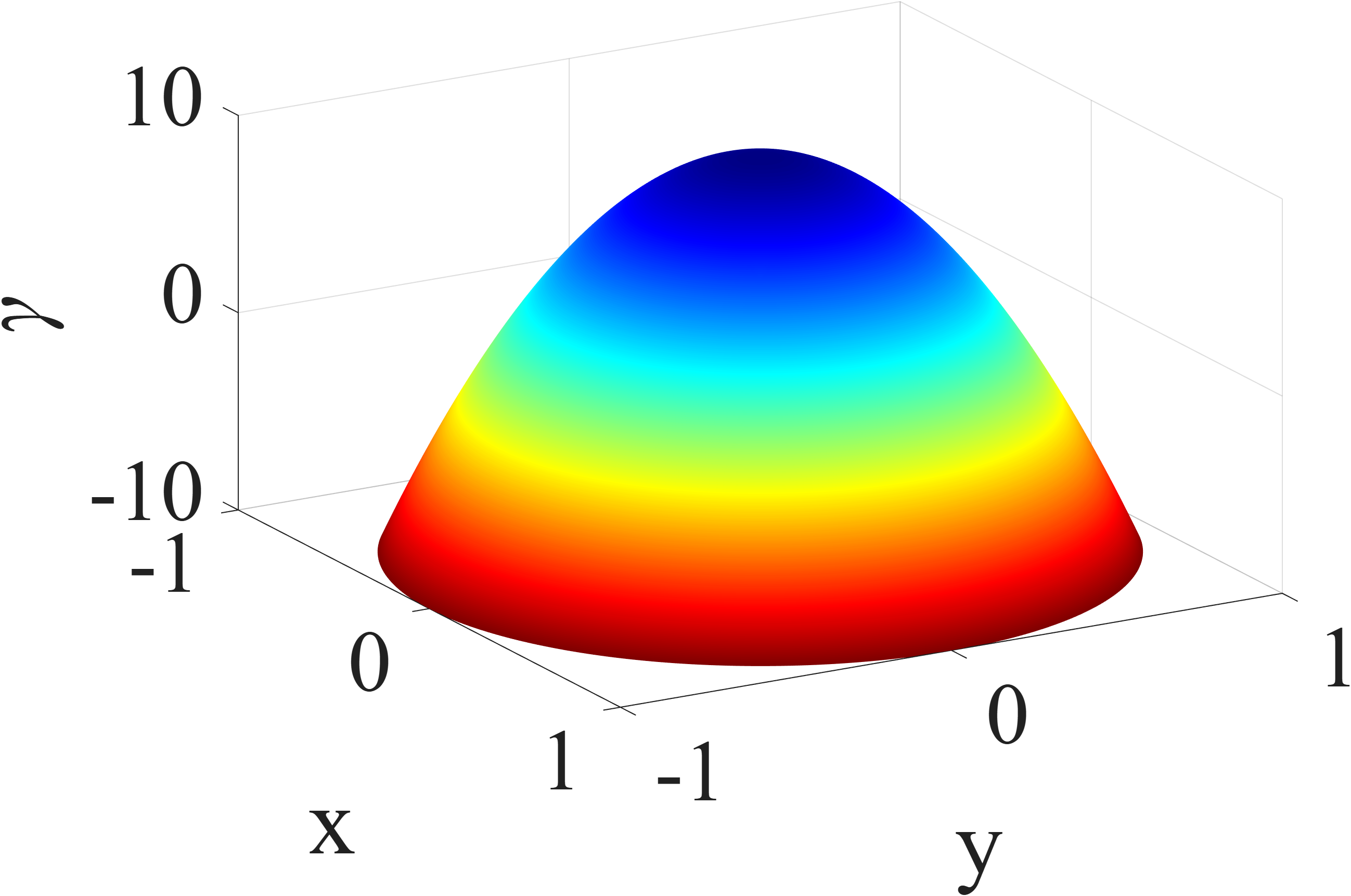}
	\end{subfigure}
	\begin{subfigure}[t]{0.325\textwidth}
		\centering
            \caption{$t = 0.0822$}
		\includegraphics[width=1\textwidth]{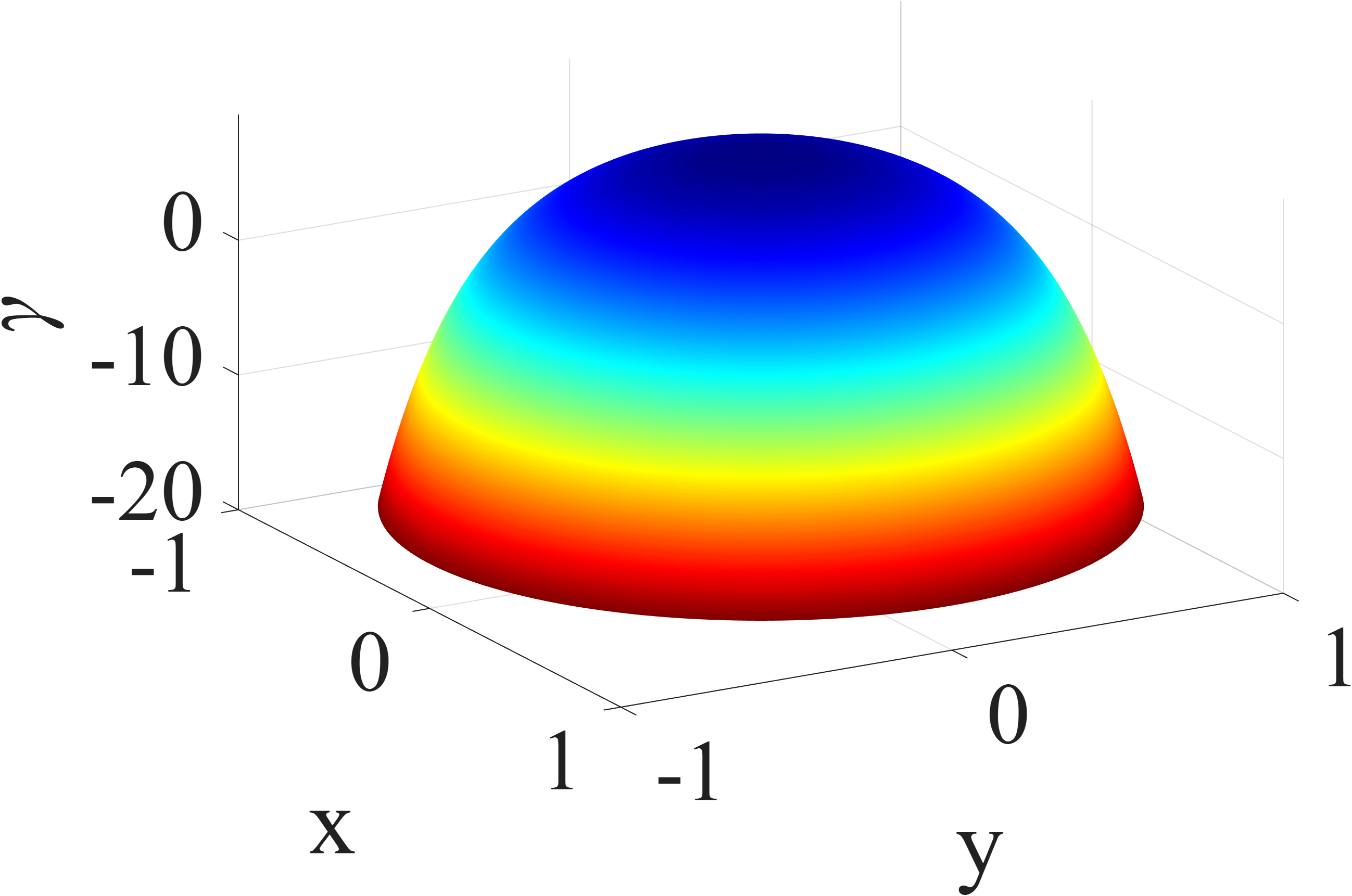}
	\end{subfigure}
	\begin{subfigure}[t]{0.325\textwidth}
		\centering
            \caption{$t = 0.1644$}
		\includegraphics[width=1\textwidth]{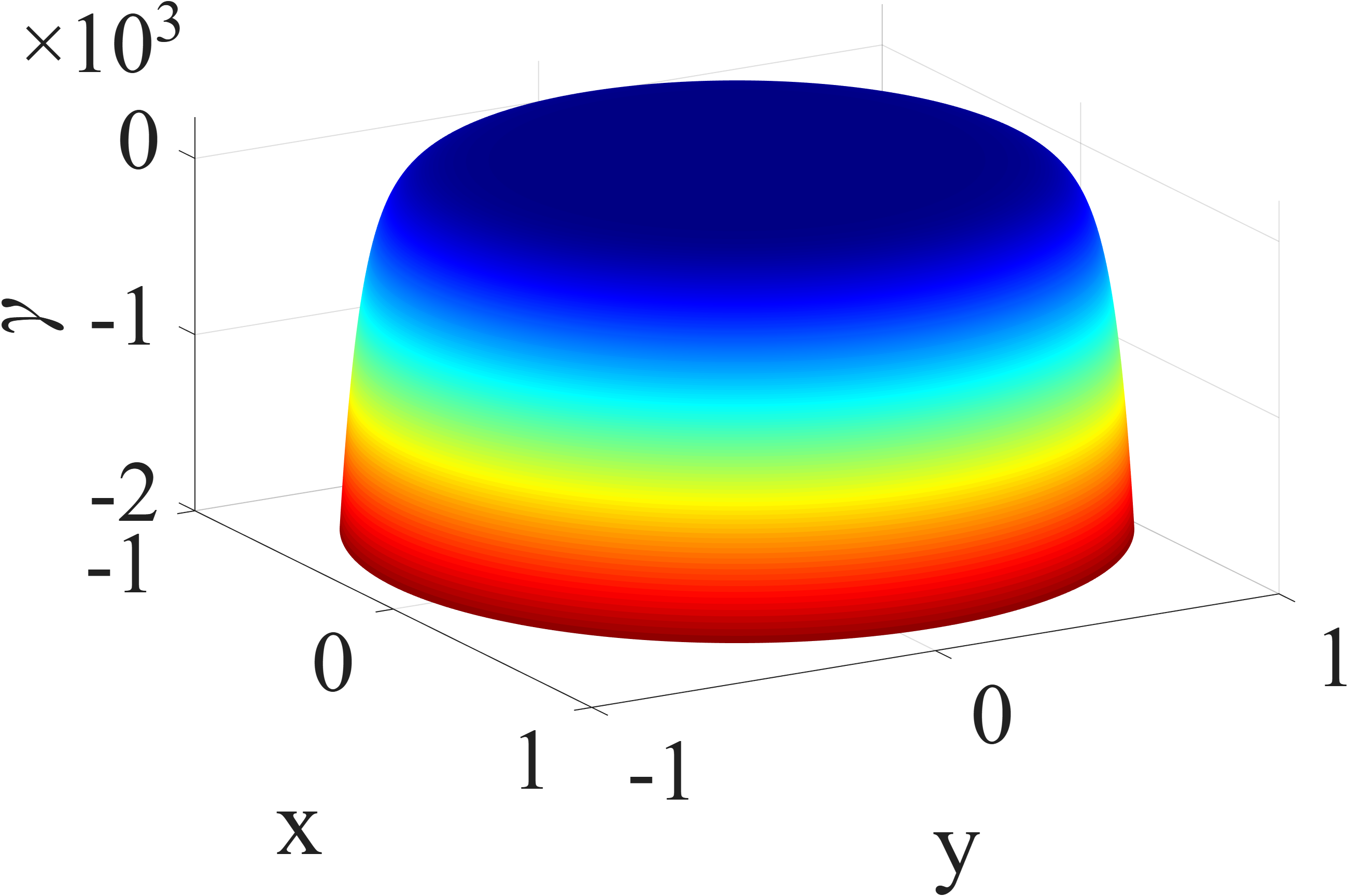}
	\end{subfigure}
        \begin{subfigure}[t]{0.325\textwidth}
		\centering
            \caption{$t = 0$}
		\includegraphics[width=1\textwidth]{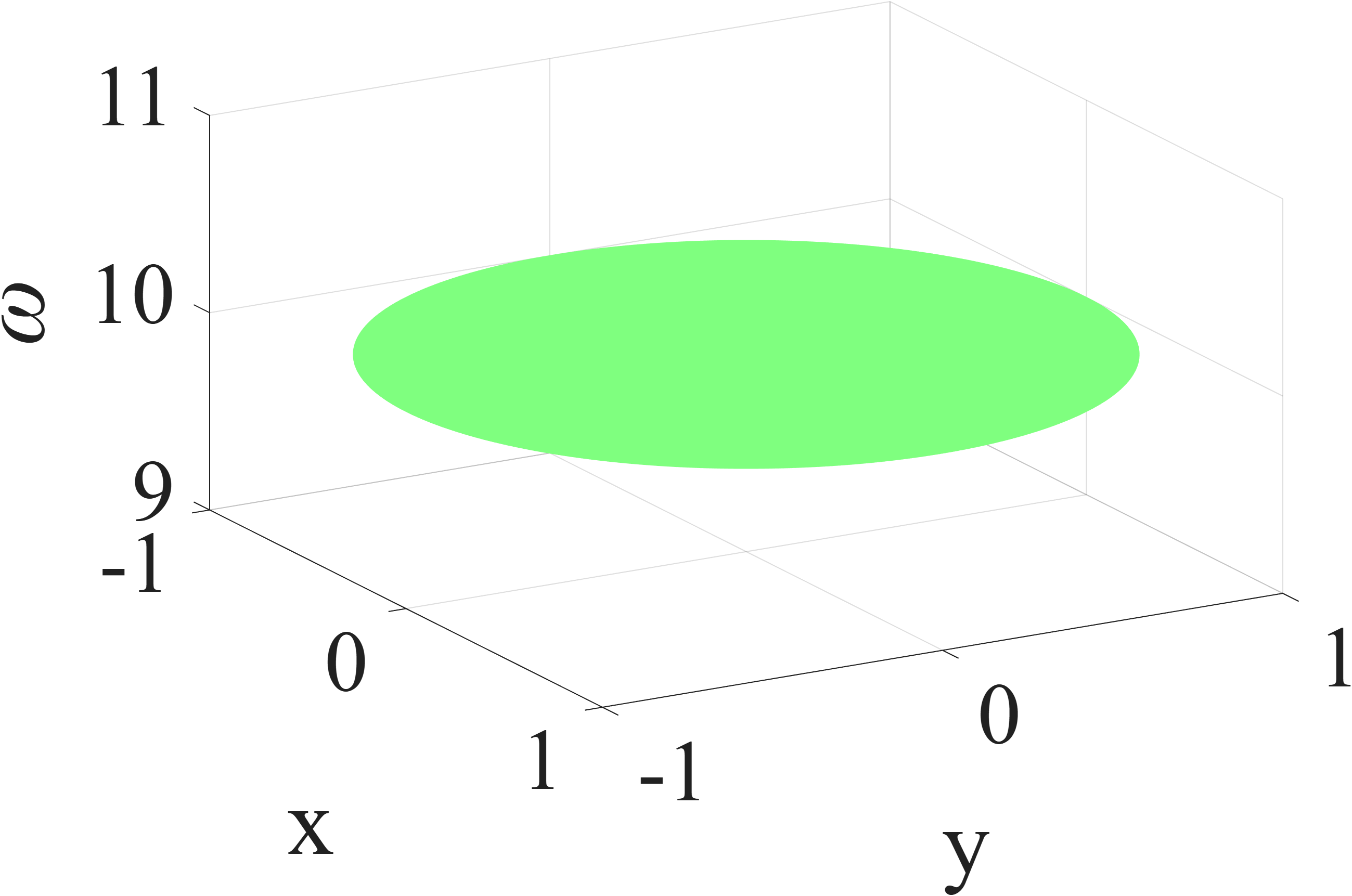}
	\end{subfigure}
        \begin{subfigure}[t]{0.325\textwidth}
		\centering
            \caption{$t = 0.0822$}
		\includegraphics[width=1\textwidth]{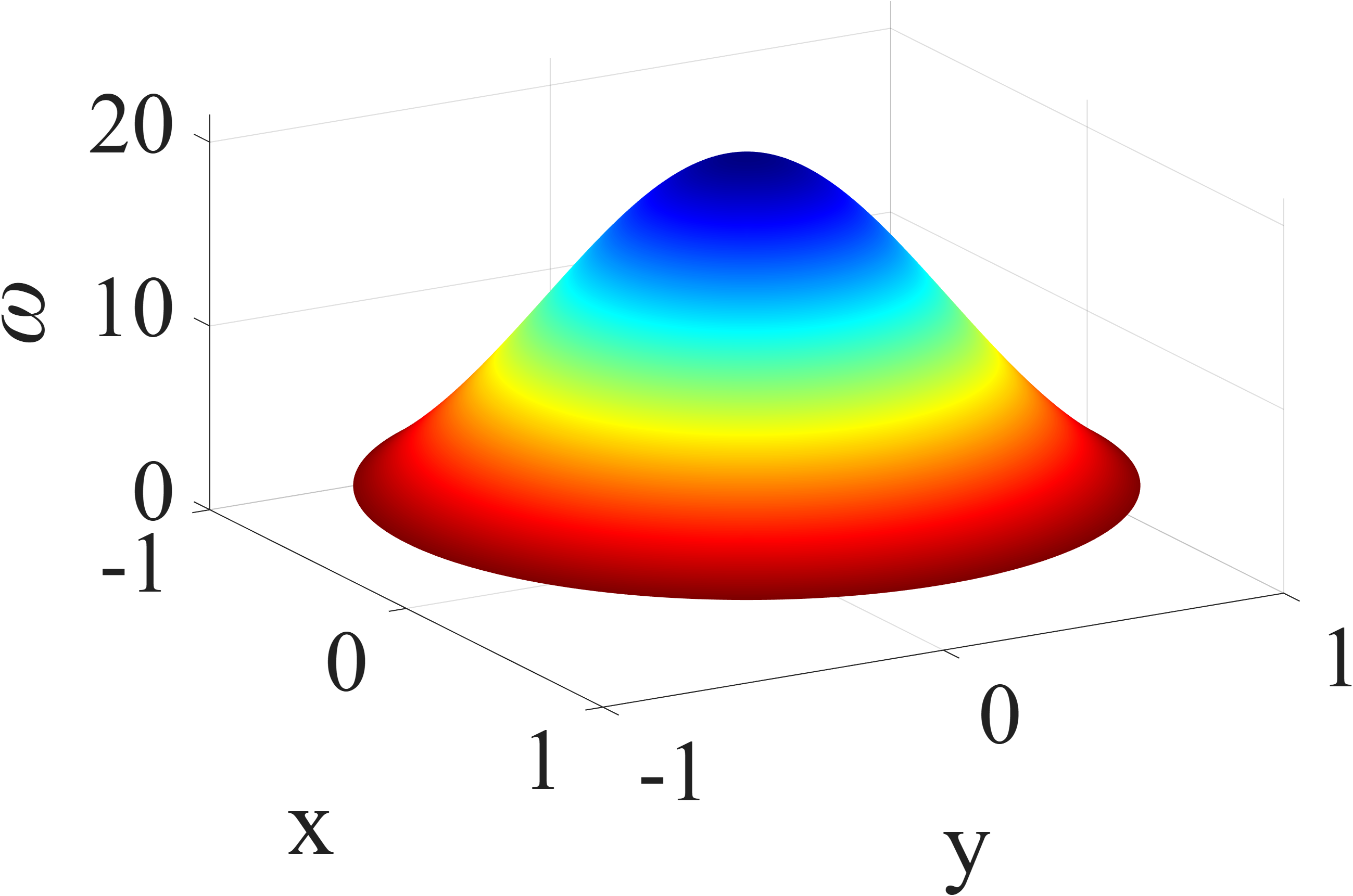}
	\end{subfigure}
        \begin{subfigure}[t]{0.325\textwidth}
		\centering
            \caption{$t = 0.1644$}
		\includegraphics[width=1\textwidth]{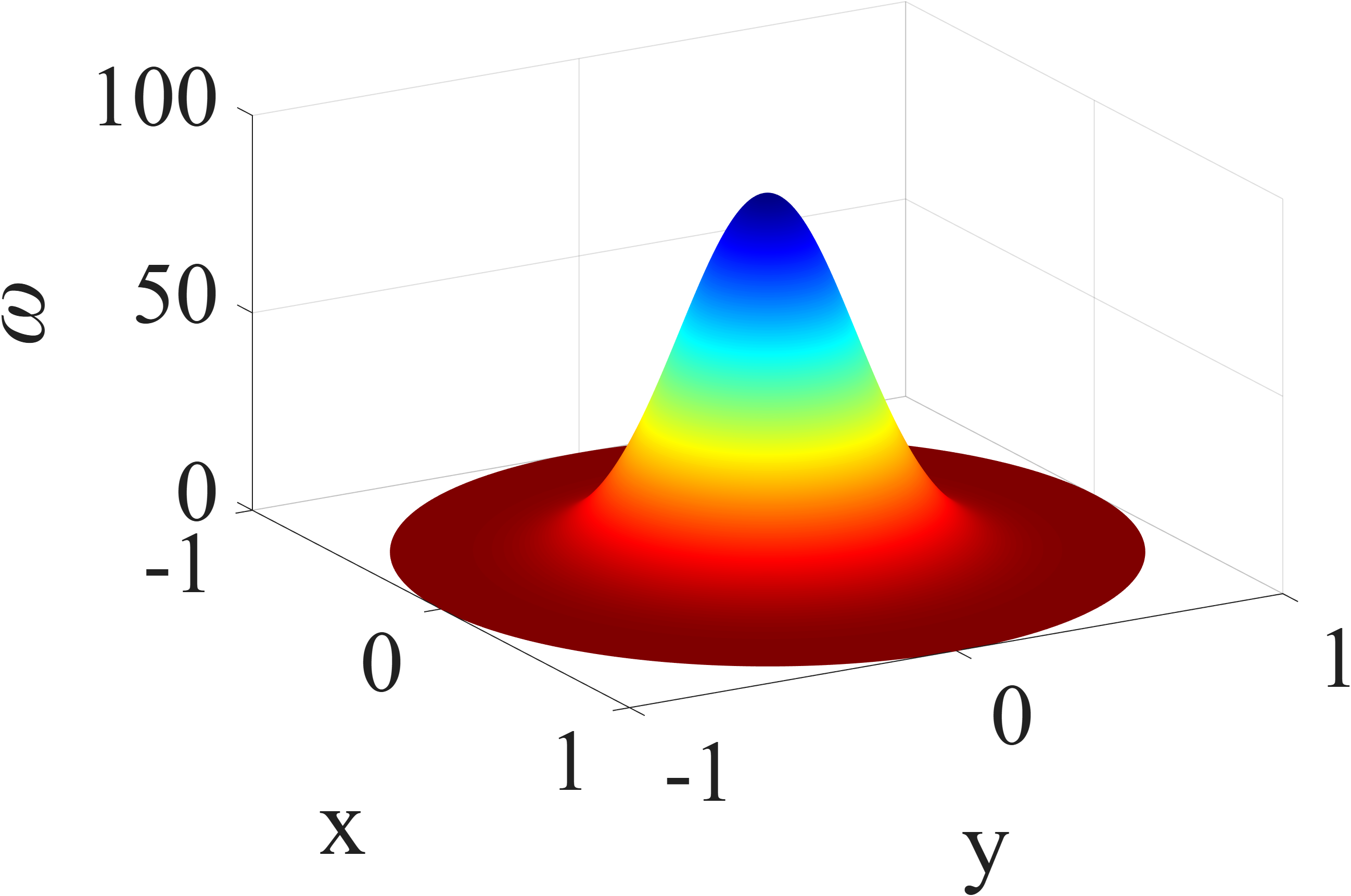}
	\end{subfigure}
\captionsetup{justification = raggedright, singlelinecheck = false}
\caption{Evolution of the parabolic stretching rate $\gamma_0(r_0) = 10 - 20r_0^2$ (hence $f=10,R=1$; {blowup is at the boundary})  {and the uniform initial plane vorticity $\omega_0(r_0) = 10$} in the Eulerian description. According to equation \eqref{eq:gamma>0}, the asymptotic behavior of the stretching rate $\gamma$ is expected to exhibit a highly skewed profile near the boundary as the solution approaches the singularity time. This characteristic is reflected in the spatial distribution observed in the figure.  {In contrast to $\gamma$'s  blowup being focused at the boundary, $\omega$ concentrates its blowup at the center ($r=0$), demonstrating the complementary nature of their singularities. The milder growth rate of $\omega$, evident from the $z$-axis scaling, is consistent with its Lambert-W-dominated asymptotics as predicted by \eqref{eq:gamma>0}.}}\label{fig:gamma->0}
\end{figure}

 {This behavior for $\gminus>0$ is demonstrated numerically in Figure~\ref{fig:gamma->0} for ${\gminus} = 10$, $R = 1$.}

These results demonstrate that the parabolic initial profile of the stretching rate can lead to finite-time singularities in both Lagrangian and Eulerian descriptions of the {Gibbon-Fokas-Doering model}. The location and nature of the blowup are highly sensitive to the sign of ${\gminus}$, despite the fact that the minimum of $\gamma_0(r_0)$ remains unchanged under the transformation $\gamma_0 \mapsto -\gamma_0$. One way to understand physically this difference in behavior is to notice that when $f<0$ the blowup occurs at a point ($r_- = 0$), whereas when $f>0$ the blowup occurs on a curve (the ring $r_-=R$). 

{
\begin{remark}
Figures \ref{fig:pathline}, \ref{fig:gamma-<0} and \ref{fig:gamma->0} were generated by using the following approach: (i) For all figures, solving the ODE for $S(t)$ in equation \eqref{ode} via a standard fourth-order Runge--Kutta method for temporal integration  with time step $\Delta t = 10^{-5}$ and with equation parameters specified in the captions of the respective figures, giving a numerical blowup time $T_*^{\mathrm{num}} = 0.1644$, close to the theoretical blowup time $T_*$ \eqref{eq:Tstar} by less than $10^{-4}$. The numerical solution was validated through a standard time-step convergence analysis, evaluating the solution at a time $t = 0.8 T_*$. (ii) For figures \ref{fig:gamma-<0} and \ref{fig:gamma->0}, using the information from point (i) above, reconstruction of $\gamma$ and $\omega$ in Eulerian coordinates via the explicit expressions in \eqref{eq:parabolaGamma}, taking a uniform initial vorticity field (so there is no need to map the pathlines), and using a regular grid of $1001$ points in the $r$-direction and $10001$ points in the $\theta$-direction. (iii) For figure \ref{fig:pathline}, using the information from point (i) above, parametric plot of the pathlines $(r,\theta,z)$ via the explicit expressions \eqref{eq:r_plot}, \eqref{eq:theta_z_plot}, with enhanced values of initial vorticity ($\omega_0 = 88$ or $\omega_0 = 380$) to better visualize the particles' rotation. The corresponding MATLAB codes are available from the corresponding author upon reasonable request.
\end{remark}
}

Furthermore, the asymptotic analysis reveals a clear self-similar character of the blowup, with both $\gamma$ and $\omega$ exhibiting power-law or logarithmic corrections governed by the Lambert $W$ function near the singularity time $T_*$. These findings motivate further investigation into the general case, where the initial stretching rate $\gamma_0(r_0)$ is not restricted to a specific functional form but is only assumed to be continuous near its minimum point.

\subsection{Initial Stretching Rate Continuous at the Minimum Point} \label{continuous}

To gain deeper insight into the relationship between the initial stretching rate and the singular behavior of the    {Gibbon-Fokas-Doering model}, we consider a more general case in which the initial stretching rate $\gamma_0(r_0)$ is assumed to be continuous in the neighborhood of its minimum point. For simplicity, we also assume that $\gamma_0(r_0)$ has only one minimum point—hence, only one potential blowup point.

Suppose there exist positive real numbers $ n_-, n_+ \in \mathbb{R}_{> 0} $ such that $ \gamma_0(r_0) $ satisfies the following asymptotic condition near the minimum point:
\begin{equation}
    \label{lambda}
    \lambda_{\pm} = \lim_{r_0 \to r_-^{\pm}} \frac{\gamma_0(r_0) - {\gminus}}{{\gminus} |r_0 - r_-|^{n_{\pm}}} \in (-\infty, 0),
\end{equation}
where $ r_- \in [0,R] $ denotes the location of the minimum point of $ \gamma_0(r_0) $, and $ {\gminus} := \gamma_0(r_-) < 0 $ is the minimum value {(notice the abuse of notation: in section} \ref{parabola} {we used $\gminus$ for either the minimum or the maximum of $\gamma_0$, whereas in this chapter $\gminus$ refers to the minimum value of $\gamma_0$, so it is necessarily negative)}. We will exclude the trivial case $ {\gminus} = 0 $ as it can only occur if $ \gamma_0(r_0) \equiv 0 $ on $[0,R]$, due to the constraint $ \langle \gamma_0 \rangle = 0 $. 

We then define the exponents $ n_- $ and $ n_+ $ as follows:
\begin{itemize}
    \item[a)] Let $ n_- $ be the positive number for which the limit $\lambda_-$ in \eqref{lambda} (from the left) is finite and negative. If, in contrast, for all $ n_- > 0 $, the limit vanishes (i.e., $ \lambda_- = 0 $), then we define $ n_- = +\infty $;
    \item[b)] Analogously, $ n_+ $ is defined as the positive number for which the limit $\lambda_+$ in \eqref{lambda} (from the right) is finite and negative. If no such $ n_+$ exists, then we set $ n_+ = +\infty $.
\end{itemize}

This definition ensures that $ n_\pm $ characterize the leading-order algebraic rate at which $ \gamma_0(r_0) $ approaches its minimum from either side. In particular, $ n_\pm < \infty $ corresponds to power-law behavior near $ r_- $, while $ n_\pm = \infty $ indicates that $ \gamma_0(r_0) $ approaches $ {\gminus} $ faster than any power law—such as in the case of Gevrey-class regularity at $ r_- $,  {(see, e.g., \cite{rodino1993linear},\cite{kukavica2011analyticity}).}

Accordingly, we can express $ \gamma_0(r_0) $ in a neighborhood of $ r_- $ via a power-law expansion:
\begin{equation}
    \label{conti3}
    \gamma_0(r_0) =
    \begin{cases}
    \displaystyle \gamma_0^{(-)}(r_0) := {\gminus} \left[1 + \lambda_{-}(r_- - r_0)^{n_-}  - (r_- - r_0)^{n_-}\varepsilon_-(r_- - r_0)\right], & r_0 \in [0, r_-], \\
    \\
    \displaystyle \gamma_0^{(+)}(r_0) := {\gminus} \left[1 + \lambda_{+}(r_0 - r_-)^{n_+}  -  (r_0 - r_-)^{n_+}\varepsilon_+(r_0 - r_-)\right], & r_0 \in [r_-, R],
    \end{cases}
\end{equation}
where the functions $ \varepsilon_\pm(x) $ are bounded in a neighborhood of $ x = 0 $ and satisfy
\begin{equation}
    \lim_{x \to 0^+} \varepsilon_\pm(x) = 0.
\end{equation}
When $ r_- = 0 $, the left branch $ \gamma_0^{(-)} $ is not applicable and $ n_- $ is not defined; similarly, when $ r_- = R $, the right branch $ \gamma_0^{(+)} $ is not applicable and $ n_+ $ is not defined.

A useful quantity in what follows is the coefficient $n$ that depends on $n_-$ and $n_+$, and dictates whether or not blowup occurs and the way it occurs. We define
$n$ by:
\begin{equation}
    \label{eq:n_def}
    n := \begin{cases}
        n_- & \text{if} \quad r_- = R\,,\\
        \max\{n_-, n_+\} & \text{if} \quad 0 < r_- < R\,,\\
        n_+ & \text{if} \quad r_- = 0\,.
    \end{cases}
\end{equation}

\begin{remark}

The cases $0< n < \infty $ imply that $ \gamma_0(r_0) $ approaches $ {\gminus} $ algebraically from the corresponding side, with leading exponent $ n$. On the other hand, $ n = \infty $ indicates super-polynomial flatness at $ r_- $. This refined definition correctly identifies the "flat" case with $ n \to \infty $.

Note that the case $ n = 0 $ is excluded by construction, as we restrict to $ n_\pm > 0 $, and the limit behavior for $ n \to 0^+ $ does not correspond to constant behavior in a meaningful asymptotic sense. Instead, a truly flat minimum would manifest as $ n = \infty $, consistent with infinite differentiability and absence of a leading power-law term.
\end{remark}

By inspecting the structure of the Lagrangian solutions for the vorticity coordinate given in equation \eqref{sol}, as well as the expression for the singularity time $T_*$ in equation \eqref{t*}, we observe that the key to analyzing these quantities lies in understanding the asymptotic behavior of the function $S(t)$. To this end, we recall the governing equations for $S(t)$ derived from \eqref{ode} and \eqref{dds}:
\begin{equation}
    \label{newode}
\begin{aligned}
    \dot{S}(t) &= \left\langle \frac{1}{1 + \gamma_0(r_0) S(t)} \right\rangle_0^{-2} = \frac{R^4}{4} \left[\int_0^R \frac{r_0}{1 + \gamma_0(r_0) S(t)} \mathrm{d}r_0 \right]^{-2}, \\
    \ddot{S}(t) &= 2 \dot{S}^{\frac{5}{2}} \left\langle \frac{\gamma_0}{(1 + \gamma_0 S)^2} \right\rangle_0 = \frac{R^8}{8} \left[\int_0^R \frac{r_0}{1 + \gamma_0(r_0) S(t)} \mathrm{d}r_0 \right]^{-5} \int_0^R \frac{\gamma_0(r_0) r_0}{\left[1 + \gamma_0(r_0) S(t)\right]^2} \mathrm{d}r_0.
\end{aligned}
\end{equation}
The evolution of $S(t)$ is governed by two dominant integrals:
\begin{equation}
    \label{inte12}
\Lambda_1(t) := \int_0^R \frac{r_0}{1 + \gamma_0(r_0) S(t)} \mathrm{d}r_0, \quad
\Lambda_2(t) := \int_0^R \frac{\gamma_0(r_0) r_0}{\left[1 + \gamma_0(r_0) S(t)\right]^2} \mathrm{d}r_0.
\end{equation}
It is important to note that $\Lambda_2(t)$ is the derivative of $\Lambda_1(t)$ with respect to $S(t)$, i.e.,
\begin{equation}
    \Lambda_2(t) = -\frac{\partial}{\partial S} \Lambda_1(t).
\end{equation}
This implies that divergence of $\Lambda_1(t)$ is sufficient—but not necessary—for $\Lambda_2(t)$ to diverge as $t \to T_*$. This distinction will be critical in determining whether finite-time singularities occur and how they behave.

{Notice that, because of the formula} \eqref{t*}{, the absence of a finite-time singularity (namely, the case $T_* = \infty$) necessarily requires $\lim_{t\to T_*}\Lambda_1(t) = \infty$.}
Therefore, a careful analysis of the convergence or divergence properties of $\Lambda_1(t)$ and $\Lambda_2(t)$ as $t \to T_*$ is essential. These integrals become improper as $S(t) \to S_* = -1/{\gminus}$, due to the vanishing denominator at the minimum point $r_0 = r_-$. In the following subsections, we examine their behavior in detail under various assumptions on $n_-$, $n_+$, and $r_-$. {We note that the presence of the max function in the definition of the coefficient $n$ in equation} \eqref{eq:n_def}{ is directly related to the fact that the original system's regularity ($T_* = \infty$) requires a divergent $\Lambda_1$, which is realized in terms of the maximum between $n_-$ and $n_+$, as we will see below in equation} \eqref{limlam2}.

\subsubsection{Convergence and Divergence of the Integrals $\Lambda_1(t)$ and $\Lambda_2(t)$} \label{3.1}

Given the identity $1 + {\gminus} S_* = 0$, the integrals $\Lambda_1(t)$ and $\Lambda_2(t)$ become improper as $t \to T_*$. This singularity arises due to the vanishing denominator at the minimum point $r_0 = r_-$, which lies within the domain $[0, R]$.

To analyze their behavior near the blowup time, we decompose each integral into a sum of improper integrals, where the essential discontinuity (i.e., the point where the integrand becomes singular) appears at the lower limit of integration. Specifically, we write:
\begin{equation}
\label{lam1}
    \begin{aligned}
    \Lambda_1(t) &= r_- \int_0^{r_-} \frac{1}{1 + \gamma_0^{(-)}(r_- - r_0) S(t)} \, \mathrm{d}r_0 + r_- \int_0^{R - r_-} \frac{1}{1 + \gamma_0^{(+)}(r_0 + r_-) S(t)} \, \mathrm{d}r_0 \\
    &\quad - \int_0^{r_-} \frac{r_0}{1 + \gamma_0^{(-)}(r_- - r_0) S(t)} \, \mathrm{d}r_0 + \int_0^{R - r_-} \frac{r_0}{1 + \gamma_0^{(+)}(r_- + r_0) S(t)} \, \mathrm{d}r_0,
\end{aligned}
\end{equation}
and
\begin{equation}
\label{lam2}
    \begin{aligned}
    \Lambda_2(t) &= r_- \int_0^{r_-} \frac{\gamma_0^{(-)}(r_- - r_0)}{[1 + \gamma_0^{(-)}(r_- - r_0) S(t)]^2} \, \mathrm{d}r_0 + r_- \int_0^{R - r_-} \frac{\gamma_0^{(+)}(r_0 + r_-)}{[1 + \gamma_0^{(+)}(r_0 + r_-) S(t)]^2} \, \mathrm{d}r_0 \\
    &\quad - \int_0^{r_-} \frac{\gamma_0^{(-)}(r_- - r_0) r_0}{[1 + \gamma_0^{(-)}(r_- - r_0) S(t)]^2} \, \mathrm{d}r_0 + \int_0^{R - r_-} \frac{\gamma_0^{(+)}(r_- + r_0) r_0}{[1 + \gamma_0^{(+)}(r_- + r_0) S(t)]^2} \, \mathrm{d}r_0.
\end{aligned}
\end{equation}
Taking the limit as $t \to T_*$, we evaluate the asymptotic behavior of these integrals at the singularity time:
\begin{equation}
\label{limlam1}
    \begin{aligned}
    \lim_{t \to T_*} \Lambda_1(t) &= -\frac{r_-}{\lambda_{-} - \varepsilon_-(\eta_0)} \lim_{\nu \to 0} \int_\nu^{r_-} r_0^{-n_-} \mathrm{d}r_0 - \frac{r_-}{\lambda_{+} - \varepsilon_+(\eta_2)} \lim_{\nu \to 0} \int_\nu^{R - r_-} r_0^{-n_+} \mathrm{d}r_0 \\
    &\quad + \frac{1}{\lambda_{-} - \varepsilon_-(\eta_1)} \lim_{\nu \to 0} \int_\nu^{r_-} r_0^{1 - n_-} \mathrm{d}r_0 - \frac{1}{\lambda_{+} - \varepsilon_+(\eta_3)} \lim_{\nu \to 0} \int_\nu^{R - r_-} r_0^{1 - n_+} \mathrm{d}r_0,
\end{aligned}
\end{equation}
and
\begin{small}
\begin{equation}
\label{limlam2}
    \begin{aligned}
    \lim_{t \to T_*} \Lambda_2(t) &= -{\gminus} \lim_{t \to T_*} \Lambda_1(t) \\
    &\quad + \frac{{\gminus} r_-}{[\lambda_{-} - \varepsilon_-(\zeta_0)]^2} \lim_{\nu \to 0} \int_\nu^{r_-} r_0^{-2n_-} \mathrm{d}r_0 + \frac{{\gminus} r_-}{[\lambda_{+} - \varepsilon_+(\zeta_2)]^2} \lim_{\nu \to 0} \int_\nu^{R - r_-} r_0^{-2n_+} \mathrm{d}r_0 \\
    &\quad - \frac{{\gminus}}{[\lambda_{-} - \varepsilon_-(\zeta_1)]^2} \lim_{\nu \to 0} \int_\nu^{r_-} r_0^{1 - 2n_-} \mathrm{d}r_0 + \frac{{\gminus}}{[\lambda_{+} - \varepsilon_+(\zeta_3)]^2} \lim_{\nu \to 0} \int_\nu^{R - r_-} r_0^{1 - 2n_+} \mathrm{d}r_0,
\end{aligned}
\end{equation}
\end{small}
where the constants $\eta_j$, $\zeta_j$ ($j = 0, 1, 2, 3$) arise from the application of the mean value theorem during the derivation.

It is important to note that modifying the upper limits of integration in equations \eqref{lam1} and \eqref{lam2} to any positive real number does not affect the convergence or divergence properties of $\Lambda_1(t)$ and $\Lambda_2(t)$. Therefore, it is permissible to set these upper limits to small positive constants. As a result, the parameters $\eta_j$ and $\zeta_j$ do not influence the convergence behavior at the singularity time, since they can be chosen arbitrarily close to zero such that the denominators in \eqref{limlam1} and \eqref{limlam2} remain nonzero.

The convergence or divergence of $\Lambda_1(t)$ and $\Lambda_2(t)$ is determined by the values of $r_-$, $n_-$, and $n_+$.  {Based on the limit as $t$ goes to $T_*$ of $\Lambda_1(t)$ and $\Lambda_2(t)$ in \eqref{limlam1} and \eqref{limlam2} under appropriate choices for $r_-$, $n_-$, and $n_+$, and using the fact that for any $\epsilon > 0, \lim_{\nu\to 0}\int_\nu^\epsilon r_0^\alpha \mathrm{d}r_0 =\infty$ when $\alpha\leq-1$, } tables \ref{tab:1} and \ref{tab:2} summarize respectively the conditions under which the integral $\Lambda_1(t)$ and $\Lambda_2(t)$ converges or diverges.  {More precisely, this} classification reveals a clear dependence on the exponents $n_-$ and $n_+$, as well as on the location of the minimum point $r_-$. In particular, in terms of the coefficient $n$ defined in equation \eqref{eq:n_def}:

\begin{itemize}
    \item[a)] When $r_- = 0$,  both integrals converge if and only if {$0 < n < 1$};
    
    \item[b)] When $r_- \neq 0$, the threshold for divergence shifts: $\Lambda_1(t)$ and $\Lambda_2(t)$ converge if and only if $0< n < \frac{1}{2}$, and diverge otherwise.
\end{itemize}

\begin{table}
  \begin{center}
\def~{\hphantom{0}}
  \begin{tabular}{|l|c|c|c|c|}
      $n \in$ & $(0,\frac{1}{2})$ & $[\frac{1}{2},1)$ & $[1,2)$ & $[2,+\infty)$ \\[7pt]
      $r_- = 0$ & Converges & Converges & Converges & Diverges \\[2pt]
      $r_- \ne 0$ & Converges & Converges & Diverges & Diverges \\
  \end{tabular}
  \caption{Convergence behavior of $\Lambda_1(t)$ as $t \to T_*$ for different values of $n$ and $r_-$ {, where $n$ is defined in \eqref{eq:n_def}}.}
  \label{tab:1}
  \end{center}
\end{table}

\begin{table}
  \begin{center}
\def~{\hphantom{0}}
  \begin{tabular}{|l|c|c|c|c|}
      $n \in$ & $(0,\frac{1}{2})$ & $[\frac{1}{2},1)$ & $[1,2)$ & $[2,+\infty)$ \\[10pt]
      $r_- = 0$ & Converges & Converges & Diverges & Diverges \\[4pt]
      $r_- \ne 0$ & Converges & Diverges & Diverges & Diverges \\
  \end{tabular}
  \caption{Convergence behavior of $\Lambda_2(t)$ as $t \to T_*$ for different values of $n$ and $r_-$ {, where $n$ is defined in \eqref{eq:n_def}}.}
  \label{tab:2}
  \end{center}
\end{table}

These results are crucial for determining whether finite-time blowup occurs in the vorticity coordinate $(\gamma, \omega)$, as both $\dot{S}(t)$ and $\ddot{S}(t)$ depend directly on $\Lambda_1(t)$ and $\Lambda_2(t)$ via the governing equations derived earlier.

\subsubsection{Asymptotic Behavior of the Integrals $\Lambda_1(t)$ and $\Lambda_2(t)$: Divergent Case} \label{divergent}

In this subsection, we focus on the asymptotic behavior of the integrals $\Lambda_1(t)$ and $\Lambda_2(t)$ when they become divergent as $t \to T_*$. By analyzing equations \eqref{lam1} and \eqref{lam2}, we observe that both $\Lambda_1(t)$ and $\Lambda_2(t)$ can be expressed as sums of improper integrals belonging to a family denoted by $I_{pq}(n_\pm; x)$, defined as:
\begin{equation}
    \label{firstipq}
I_{pq}(n_\pm; x) := {\int}_0^{x} \frac{\left[\gamma_0^{(\pm)}(r_- \pm r_0)\right]^{q-1} r_0^p}{\left[1 + \gamma_0^{(\pm)}(r_- \pm r_0) S(t)\right]^q} \mathrm{d}r_0,
\end{equation}
where $p \geq 0$ and $q \geq 1$ are integers. Using this definition, the integrals $\Lambda_1(t)$ and $\Lambda_2(t)$  {expressed in \eqref{lam1} and \eqref{lam2}} can be rewritten in terms of these components as follows:
\begin{equation}
    \label{ilam}
\Lambda_j(t) = r_- \left[ I_{0j}(n_+; R - r_-) + I_{0j}(n_-; r_-) \right] + \left[ I_{1j}(n_+; R - r_-) - I_{1j}(n_-; r_-) \right], \quad j = 1, 2.
\end{equation}
Based on this decomposition, we derive the asymptotic behavior of $\Lambda_1(t)$ and $\Lambda_2(t)$ under different assumptions on the exponents $n_-$, $n_+$, and the location of the minimum point $r_-$.\\

\begin{itemize}
    \item[\textbf{Case 1.}] $n_-, n_+ \in \mathbb{R}_+$ when they are defined, namely irrespective of the value of $r_- \in [0,R]$. 

Substituting the expression for $\gamma_0^{(\pm)}$ into equation \eqref{firstipq}, we obtain the leading-order asymptotic form:
\begin{equation}
    \begin{aligned}
    I_{pq}(n_\pm; x) &= \int_0^{x} \frac{\left[\gamma_0^{(\pm)}(r_- \pm r_0)\right]^{q-1} r_0^p}{\left[1 + \gamma_0^{(\pm)}(r_- \pm r_0) S(t)\right]^q} \mathrm{d}r_0 \\
    &\sim \int_0^{x} \frac{\left[1 + (\lambda_{\pm} - \varepsilon_\pm) r_0^{n_\pm} \right]^{q-1} {\gminus}^{q-1} r_0^p}{\left[1 - \frac{S}{S_*} - (\lambda_{\pm} - \varepsilon_\pm) r_0^{n_\pm} \right]^q} \mathrm{d}r_0 \\
    &\sim (-1)^q \frac{{\gminus}^{q-1}}{\lambda_{\pm}^q} \int_0^x \frac{r_0^p}{\left(r_0^{n_\pm} + B_\pm \right)^q} \mathrm{d}r_0 \\
    &\sim (-1)^q \frac{{\gminus}^{q-1}}{\lambda_{\pm}^q} \cdot \frac{x^{p+1}}{(p+1)B_\pm^q} {}_2F_1\left(q, \frac{1+p}{n_\pm}, 1+\frac{1+p}{n_\pm}; -\frac{x^{n_\pm}}{B_\pm}\right),
\end{aligned}
\end{equation}
where $B_\pm = \frac{{\gminus}}{\lambda_{\pm}}(S_* - S)$, and ${}_2F_1$ denotes the Gaussian hypergeometric function.

{To continue the derivation, we apply the series expansion of the Gaussian hypergeometric function ${}_2F_1(a,b;c;z)$ into equation (3.39), and find the leading-order term in terms of $B_{\pm}\rightarrow 0$. Notice that $\left|-\frac{x^{n_{\pm}}}{B_\pm}\right|\rightarrow\infty$ in this case. Thus, we need to select the series expansion of ${}_2F_1(a,b;c;z)$ at the branch $|z|>1$, and the special case $n_\pm = \frac{1+p}{q}$ corresponds to the logarithmic leading-order term. Once this is noticed}, we obtain the following asymptotic forms:
\begin{equation}
    \label{ipq}
I_{pq}(n_\pm; x) \sim
\begin{cases}
\displaystyle (-1)^q \frac{{\gminus}^{q-1}}{\lambda_{\pm}^q} \cdot \frac{\Gamma\left(q - \frac{1+p}{n_\pm}\right)\Gamma\left(1 + \frac{1+p}{n_\pm}\right)}{(p+1)\Gamma(q)} B_\pm^{\frac{1+p}{n_\pm} - q}, & n_\pm > \frac{1+p}{q}, \\
\\
\displaystyle (-1)^q \frac{q! {\gminus}^{q-1}}{(1+p)\lambda_{\pm}^q} \ln B_\pm, & n_\pm = \frac{1+p}{q},
\end{cases}
\end{equation}
as $t \to T_*$, where $B_\pm \to 0$. The integral $I_{pq}(n_\pm; x)$ converges when $n_\pm < \frac{1+p}{q}$, which is why this case does not appear in the above classification.

It is worth noting that the asymptotics of ${}_2F_1$ differ depending on whether $\frac{1+p}{n_\pm} - q$ is an integer or not. These two cases will be considered separately in the detailed analysis.

By substituting the asymptotic results from \eqref{ipq} into equation \eqref{ilam} and simplifying, we obtain the leading-order behavior of $\Lambda_1(t)$ and $\Lambda_2(t)$ near the singularity time.
    \begin{itemize}
        \item[\textbf{Subcase 1a.}] Case $r_- = 0$: 
        
For $r_- = 0$, only the right-side integral contributes. The asymptotic behavior is given by:
\begin{small}
\begin{equation}
    \Lambda_1(t) \sim
\begin{cases}
\displaystyle -\frac{\pi}{n_+ \lambda_{+} \sin\left(\frac{2\pi}{n_+}\right)} B_+^{\frac{2}{n_+} - 1}, & n_+ > 2, \\
\\
\displaystyle \frac{1}{2\lambda_+} \ln B_+, & n_+ = 2,
\end{cases},\quad
\Lambda_2(t) \sim
\begin{cases}
\displaystyle \frac{(n_+ - 2){\gminus} \pi}{\lambda_{+}^2 n_+^2 \sin\left(\frac{2\pi}{n_+}\right)} B_+^{\frac{2}{n_+} - 2}, & n_+ > 1, \\
\\
\displaystyle -\frac{{\gminus}}{\lambda_+^2} \ln B_+, & n_+ = 1.
\end{cases}.
\end{equation}
\end{small}

        \item[\textbf{Subcase 1b.}] Case $r_- \neq 0$:

Let $n$ be as defined in equation \eqref{eq:n_def}. Then we have:
\begin{equation}
    \Lambda_1(t) \sim
\begin{cases}
\displaystyle -\frac{\pi r_- \delta}{n \lambda_n \sin\left(\frac{\pi}{n}\right)} B^{\frac{1}{n} - 1}, & n > 1, \\
\\
\displaystyle \frac{r_- \delta}{\lambda_1} \ln B, & n = 1,
\end{cases},\quad
\Lambda_2(t) \sim
\begin{cases}
\displaystyle \frac{(n - 1){\gminus} \pi r_- \delta}{\lambda_n^2 n^2 \sin\left(\frac{\pi}{n}\right)} B^{\frac{1}{n} - 2}, & n > \frac{1}{2}, \\
\\
\displaystyle -\frac{2 r_- {\gminus} \delta}{\lambda_{0.5}^2} \ln B, & n = \frac{1}{2},
\end{cases}
\end{equation}
where $\lambda_n := \lambda_\pm$ if $n_\pm = n$. Here, the coefficient $\delta$ accounts for the symmetry between $n_-$ and $n_+$, and is defined as:
\begin{equation}
    \label{delta}
{\delta} =
\begin{cases}
1, & n_- \neq n_+, \\
2, & n_- = n_+.
\end{cases}
\end{equation}
    \end{itemize}
    
    \item[\textbf{Case 2.}] {$n_-n_+ = +\infty$ when they are defined, namely irrespective of the value of $r_- \in [0,R]$.}

To analyze this scenario, we examine the limit when $n_\pm \to \infty$ of  $I_{pq}(n_\pm;x)$, based on the fact that $\lim_{n_\pm\to \infty} \gamma_0^{(\infty)}(r_- \pm r_0)  = f$:
\begin{equation}
    \begin{aligned}
\lim_{n_\pm\to \infty} I_{pq}(n_\pm; x) & = \lim_{n_\pm\to \infty} \int_0^{x} \frac{\left[\gamma_0^{(\infty)}(r_- \pm r_0)\right]^{q-1} r_0^p}{\left[1 + \gamma_0^{(\infty)}(r_- \pm r_0) S(t)\right]^q} \mathrm{d}r_0\\[5pt] &\sim \frac{1}{{\gminus}} \int_0^{x} \frac{(-1)^q r_0^p}{(S_* - S)^q} \mathrm{d}r_0\\[8pt] &\sim \frac{(-1)^q x^{p+1}}{(p+1){\gminus}} (S_* - S)^{-q}.
\end{aligned}
\end{equation}
Substituting this result together with \eqref{ipq} into equation \eqref{ilam}, we find the asymptotic forms of $\Lambda_1(t)$ and $\Lambda_2(t)$ as $t \to T_*$:
\begin{equation}
\label{aai}
    \Lambda_1(t) \sim
\begin{cases}
\displaystyle -\frac{r_-^2}{2{\gminus}}(S_* - S)^{-1}, & r_- \neq 0,\quad n_- = +\infty,\quad n_+ \in \mathbb{R}_+, \\
\\
\displaystyle \frac{r_-^2 - R^2}{2{\gminus}}(S_* - S)^{-1}, & r_- \neq R,\quad n_- \in \mathbb{R}_+,\quad n_+ = +\infty, \\
\\
\displaystyle -\frac{R^2}{2{\gminus}}(S_* - S)^{-1}, & r_- \in [0,R],\quad n_- = n_+ = +\infty,
\end{cases}
\end{equation}
and
\begin{equation}
\label{aaj}
    \Lambda_2(t) \sim
\begin{cases}
\displaystyle -\frac{1}{2} r_-^2(S_* - S)^{-2}, & r_- \neq 0,\quad n_- = +\infty,\quad n_+ \in \mathbb{R}_+, \\
\\
\displaystyle \frac{1}{2}(r_-^2 - R^2)(S_* - S)^{-2}, & r_- \neq R,\quad n_- \in \mathbb{R}_+,\quad n_+ = +\infty, \\
\\
\displaystyle -\frac{1}{2} R^2(S_* - S)^{-2}, & r_- \in [0,R],\quad n_- = n_+ = +\infty.
\end{cases}
\end{equation}
These results confirm that the divergence properties of $\Lambda_1(t)$ and $\Lambda_2(t)$ depend crucially on the values of $n_-$, $n_+$, and the location of the minimum point $r_-$. This classification provides the foundation for determining whether finite-time blowup occurs and how it behaves asymptotically.
    
\end{itemize}

\subsubsection{Asymptotic Behavior of the Integrals $\Lambda_1(t)$ and $\Lambda_2(t)$: Convergent Case} \label{3.3}

In this subsection, we consider the case where the integrals $\Lambda_1(t)$ and $\Lambda_2(t)$ remain finite as $t \to T_*$. In such situations, there is no need to decompose them into multiple improper components as was done in equations \eqref{lam1} and \eqref{lam2}. We define:
\begin{equation}
 \Lambda_{j*}:=    \lim_{t \to T_*} \Lambda_j(t), \quad j = 1, 2, 3,
\end{equation}
and analyze their asymptotic behavior near the singularity time.

Starting from the definition of $\Lambda_1(t)$, we expand it as follows:
\begin{equation}
\label{llam1}
    \begin{aligned}
\Lambda_1(t) &= \int_0^R \frac{r_0}{1 + \gamma_0(r_0) S(t)} \mathrm{d}r_0 \\
&= \int_0^R \frac{r_0}{1 + \gamma_0(r_0) S_*} \mathrm{d}r_0 + (S_* - S) \int_0^R \frac{\gamma_0(r_0) r_0}{(1 + \gamma_0(r_0) S)(1 + \gamma_0(r_0) S_*)} \mathrm{d}r_0 \\
&= \Lambda_{1*} + \frac{\lambda_n B}{{\gminus}} \int_0^R \frac{\gamma_0(r_0) r_0}{(1 + \gamma_0(r_0) S)(1 + \gamma_0(r_0) S_*)} \mathrm{d}r_0,
\end{aligned}
\end{equation}
where $B = \frac{{\gminus}}{\lambda_n}(S_* - S)$ tends to zero as $t \to T_*$, and $\Lambda_{1*}$ denotes the limiting value of $\Lambda_1(t)$ at the singularity time.

Similarly, for $\Lambda_2(t)$, we obtain:
\begin{equation}
\label{llam2}
\begin{aligned}
\Lambda_2(t) &= \int_0^R \frac{\gamma_0(r_0) r_0}{[1 + \gamma_0(r_0) S(t)]^2} \mathrm{d}r_0 = \Lambda_{2*} + \frac{\lambda_n B}{{\gminus}} \int_0^R \frac{[2 + \gamma_0(S_* + S)] \gamma_0^2 r_0}{(1 + \gamma_0 S)^2 (1 + \gamma_0 S_*)^2} \mathrm{d}r_0.
\end{aligned}
\end{equation}
It is important to note that the second term in equation \eqref{llam1} has already been analyzed in Section~\ref{divergent}, and its limit as $t \to T_*$ can be directly used to derive the asymptotic behavior of $\Lambda_1(t)$.

The key challenge lies in analyzing the integral appearing in the second term of \eqref{llam2}, particularly whether it diverges as $t \to T_*$. To this end, we observe that:
\begin{equation}
    \lim_{t \to T_*} \int_0^R \frac{[2 + \gamma_0(S_* + S)] \gamma_0^2 r_0}{(1 + \gamma_0 S)^2 (1 + \gamma_0 S_*)^2} \mathrm{d}r_0 = \lim_{t \to T_*} \int_0^R \frac{2 \gamma_0^2 r_0}{(1 + \gamma_0 S)^3} \mathrm{d}r_0 =: 2 \lim_{t \to T_*} \Lambda_3(t).
\end{equation}
We can express $\Lambda_3(t)$ in terms of the family of integrals $I_{pq}(n_\pm; x)$ defined earlier:
\begin{equation}
    2 \Lambda_3(t) = 2 r_- \left[ I_{03}(n_+; R - r_-) + I_{03}(n_-; r_-) \right] + 2 \left[ I_{13}(n_+; R - r_-) - I_{13}(n_-; r_-) \right].
\end{equation}
{Applying the asymptotics of $I_{pq}$-type integral in equation \eqref{ipq} and extracting the leading-order term}, we derive the following asymptotic forms:
\begin{equation}
    2 \Lambda_3(t) \sim
\begin{cases}
\displaystyle \frac{\Gamma\left(3 - \frac{1}{n}\right)\Gamma\left(1 + \frac{1}{n}\right) {\gminus}^2 r_- {\delta}}{\lambda_n^3} B^{\frac{1}{n} - 3}, & r_- \neq 0,\quad n \in \left(\frac{1}{3}, \frac{1}{2}\right), \\
\\
\displaystyle - \frac{6 {\gminus}^2 r_- {\delta}}{\lambda_n^3} \ln B, & r_- \neq 0,\quad n = \frac{1}{3}, \\
\\
\displaystyle \frac{2 {\gminus}^2 {\delta}}{\lambda_n^3} \cdot \frac{(1 - n)(2 - n)\pi}{n^3 \sin\left(\frac{2\pi}{n}\right)} B^{\frac{2}{n} - 3}, & r_- = 0,\quad n \in \left(\frac{2}{3}, 1\right), \\
\\
\displaystyle - \frac{3 {\gminus}^2 {\delta}}{\lambda_n^3} \ln B, & r_- = 0,\quad n = \frac{2}{3}, \\
\\
\displaystyle 2 \Lambda_{3*} + o(1), &
\begin{cases}
r_- \neq 0,\quad n \in \left(0, \frac{1}{3}\right), \\
r_- = 0,\quad n \in \left(0, \frac{2}{3}\right),
\end{cases}
\end{cases}
\end{equation}
where $n$ is defined in equation \eqref{eq:n_def}, and the coefficient ${\delta}$ is defined in equation \eqref{delta}.

Based on these results, we summarize the asymptotic behaviors of $\Lambda_1(t)$ and $\Lambda_2(t)$ in tables \ref{table:lambda1} and \ref{table:lambda2}, where the constants $G_n$ and $L_j$ are defined as:
\begin{equation}
\label{ll}
    G_n = \Gamma\left(3 - \frac{1}{n}\right)\Gamma\left(1 + \frac{1}{n}\right), \quad L_j = \frac{\pi j}{n \sin\left(\frac{\pi j}{n}\right)}, \quad j = 1, 2.
\end{equation}
The left portion of the table corresponds to the case $r_- = 0$, while the right portion applies to $r_- \in (0, R]$. These results, together with those derived in equations \eqref{aai} and \eqref{aaj} for the special cases where $n_- n_+ = \infty$, allow us to determine the structure of singular solutions in the    {Gibbon-Fokas-Doering model} model.

\begin{table}
  \begin{center}
\def~{\hphantom{0}}
  \begin{tabular}{|c|c|c|}
      \textbf{Interval of $n$}  & $\Lambda_1(t)$ & $\Lambda_2(t)$ \\[12pt]
      $\left(0,\frac{2}{3}\right)$ & $\Lambda_{1*} + \dfrac{\lambda_n\Lambda_{2*}}{f} B$ & $\Lambda_{2*} + \dfrac{2\lambda_n\Lambda_{3*}}{f} B$ \\[12pt]
      $\left\{\frac{2}{3}\right\}$ & Same as above & $\Lambda_{2*} + \dfrac{3f}{\lambda_n^2} B \ln B$ \\[12pt]
      $\left(\frac{2}{3},1\right)$ & Same as above & $\Lambda_{2*} + \dfrac{(1-n)(2-n)fL_2}{\lambda_n^2 n^2} B^{\frac{2}{n}-2}$ \\[12pt]
      $\{1\}$ & $\Lambda_{1*} - \dfrac{1}{\lambda_1} B \ln B$ & $-\dfrac{f}{\lambda_1} \ln B$ \\[12pt]
      $(1,2)$ & $\Lambda_{1*} + \dfrac{(n-2)L_2}{2n\lambda_n} B^{\frac{2}{n}-1}$ & $\dfrac{(n-2)fL_2}{2n\lambda_n^2} B^{\frac{2}{n}-2}$ \\[12pt]
      $\{2\}$ & $\dfrac{1}{2\lambda_2} \ln B$ & Same as above \\[12pt]
      $(2,+\infty)$ & $-\dfrac{L_2}{2\lambda_n} B^{\frac{2}{n}-1}$ & Same as above \\
  \end{tabular}
  \caption{Asymptotic expressions for $\Lambda_1(t)$ and $\Lambda_2(t)$ when $r_- = 0$.}
  \label{table:lambda1}
  \end{center}
\end{table}

\begin{table}
  \begin{center}
\def~{\hphantom{0}}
  \begin{tabular}{|c|c|c|}
      \textbf{Interval of $n$}  & $\Lambda_1(t)$ & $\Lambda_2(t)$ \\[12pt]
      $\left(0,\frac{1}{3}\right)$ & $\Lambda_{1*} + \dfrac{\lambda_n\Lambda_{2*}}{f} B$ & $\Lambda_{2*} + \dfrac{2\lambda_n\Lambda_{3*}}{f} B$ \\[12pt]
      $\left\{\frac{1}{3}\right\}$ & Same as above & $\Lambda_{2*} - \dfrac{6f r_- \delta}{\lambda_n^2} B \ln B$ \\[12pt]
      $\left(\frac{1}{3},\frac{1}{2}\right)$ & Same as above & $\Lambda_{2*} + \dfrac{G_n f r_- \delta}{\lambda_n^2} B^{\frac{1}{n}-2}$ \\[12pt]
      $\left\{\frac{1}{2}\right\}$ & $\Lambda_{1*} - \dfrac{2r_- \delta}{\lambda_{0.5}} B \ln B$ & $-\dfrac{2f r_- \delta}{\lambda_{0.5}^2} \ln B$ \\[12pt]
      $\left(\frac{1}{2},1\right)$ & $\Lambda_{1*} + \dfrac{(n-1)r_- \delta L_1}{n\lambda_n} B^{\frac{1}{n}-1}$ & $\dfrac{(n-1)f r_- \delta L_1}{n\lambda_n^2} B^{\frac{1}{n}-2}$ \\[12pt]
      $\{1\}$ & $\dfrac{r_- \delta}{\lambda_1} \ln B$ & Same as above \\[12pt]
      $(1,+\infty)$ & $-\dfrac{r_- \delta L_1}{\lambda_n} B^{\frac{1}{n}-1}$ & Same as above \\
  \end{tabular}
  \caption{Asymptotic expressions for $\Lambda_1(t)$ and $\Lambda_2(t)$ when $r_- \in (0,R]$.}
  \label{table:lambda2}
  \end{center}
\end{table}

This classification provides a comprehensive understanding of how the integrals behave when they converge as $t \to T_*$. Notably, the convergence or divergence depends heavily on the exponents $n_-$ and $n_+$, as well as the location of the minimum point $r_-$.

From this analysis, we see that, regarding the asymptotic behavior of $\Lambda_1$ and $\Lambda_2$ as $t\to T_*$:

\begin{itemize}
    \item For small values of $n_{\pm}$, the integrals remain close to their limiting values $\Lambda_{1*}$ and $\Lambda_{2*}$, indicating {convergent behavior};
    
    \item As $n$ {increases}, {divergent behavior appears} and becomes more pronounced;
    
    \item The presence of logarithmic terms suggests a slower {divergence}, whereas power-law terms indicates a faster divergence or convergence, depending on the value of $n$.

\end{itemize}

These asymptotic results will serve as the foundation for determining the regularity or blowup nature of the vorticity coordinate $(\gamma, \omega)$ and the vertical pathline $z(z_0, r_0, t)$ in the subsequent analysis. {Recall that a convergent behavior of $\Lambda_1$ as $t\to T_*$ implies a finite-time singularity of the solution to the original fluid equations (i.e. $T_*<\infty$).}

 \subsubsection{Asymptotic Behaviors of the Vorticity Coordinate}

Having established the properties and limiting values of the integrals $\Lambda_1(t)$ and $\Lambda_2(t)$ in Sections~\ref{3.1}–\ref{3.3}, we now proceed to derive the asymptotic behavior of the vorticity coordinate $(\omega, \gamma)$ and the vertical pathline component $z$ as $t \to T_*$. The time-dependent functions $\dot{S}(t)$ and $\ddot{S}(t)$ are obtained from the governing equations in \eqref{newode} and take the following form:
\begin{equation}
\label{eq:TwithLambda}
    \dot{S}(t) = \frac{R^4}{4\Lambda_1^2(t)}, \quad
\ddot{S}(t) = \frac{R^8 \Lambda_2(t)}{8\Lambda_1^5(t)}, \quad
T_* = \frac{4}{R^4} \int_0^{S_*} \Lambda_1^2(t)\, dS.
\end{equation}
Using these expressions, the components of the vorticity coordinate and the vertical position can be rewritten near the singularity time as:

\begin{equation}
    \begin{aligned}
    \gamma(r(r_-,t), t) &\sim -\frac{R^4}{4\Lambda_1^2(t)} (S_* - S)^{-1} - \frac{R^4 \Lambda_2(t)}{4\Lambda_1^3(t)}, \\
    \omega(r(r_-,t), t) &\sim {\gminus} \omega_0(r_-) \frac{2\Lambda_1(t)}{R^2} (S_* - S), \\
    z(z_0, r(r_-,t), t) &\sim {\gminus} z_0 \frac{2\Lambda_1(t)}{R^2} (S_* - S).
\end{aligned}
\end{equation}
Based on the asymptotics of $\Lambda_1(t)$ and $\Lambda_2(t)$ derived earlier, we classify the blowup behaviors into three distinct cases, depending on the structure of the initial stretching rate $\gamma_0(r_0)$, again referring to the coefficient $n$ defined in equation \eqref{eq:n_def}:

\begin{itemize}
    \item[a)] \textbf{Flat minimum:} when $n = +\infty$, i.e., at least one side of $\gamma_0(r_0)$ is flat at the minimum point;
    
    \item[b)] \textbf{Central minimum:} when $n$ is finite and the minimum point lies at the center: $r_- = 0$;

    \item[c)] \textbf{Non-central minimum:} when $n$ is finite and the minimum point lies away from the center: $r_- \in (0, R]$.
\end{itemize}

To simplify the notation and streamline the presentation of asymptotic results, we define the following auxiliary variables: 
\begin{equation}
    \label{kl}
\begin{aligned}
K_{j^{\pm}} &= \frac{n(2j - n)\lambda_{\pm} {\gminus} R^4 \sin^2\left(\frac{\pi j}{n}\right)}{4\pi^2 (r_- {\delta})^{4 - 2j}}, \quad
    L_j = \frac{\pi j}{n \sin\left(\frac{\pi j}{n}\right)}, \\
    G_n &= \Gamma\left(3 - \frac{1}{n}\right) \Gamma\left(1 + \frac{1}{n}\right), \quad
    \Delta_{jt^{\pm}} = \sqrt{\lambda_{\pm} {\gminus} (T_* - t)}, \quad j = 1, 2.
\end{aligned}
\end{equation}
These quantities appear frequently in the asymptotic expressions for $\omega$, $\gamma$, and $z$, and their definitions encapsulate the dependence on the cylinder radius $R$, the location of the blowup point $r_-$, and the exponents $n_-, n_+$ that characterize the local behavior of $\gamma_0(r_0)$ near its minimum.

The subsequent analysis will focus on how the vorticity coordinate and vertical pathline behave near the singularity time under these different scenarios. In particular, we examine the blowup scaling laws and determine whether the divergence occurs in finite time or not.\\

 {\noindent \textbf{a) Flat minimum.}} Firstly, table \ref{table2} summarizes the case of \textbf{flat minimum}, where $n_- n_+ = +\infty$, meaning that at least one side (either left or right) of the blowup point remains flat within a neighborhood. The result showing an infinite singularity time $T_* = \infty$ indicates that the    {Gibbon-Fokas-Doering model} remains regular for all finite times. Consequently, it is not meaningful to discuss the asymptotic behavior of the flow near the singularity in this scenario.

{This observation implies a key physical insight: \textit{the flatter the initial stretching rate $\gamma_0(r_0)$ near its minimum point, the less likely it is for a finite-time singularity to occur in the Lagrangian description of the flow.} {In more quantitative terms, in the case $r_-=0$ (singularity at the center of the cylinder), an initial profile $\gamma_0(r_0) \sim r_0^{n}$ near the center will look `flatter' if $n(>0)$ is chosen larger: for $n\geq 4$ there is `grow-up' (the singularity time is equal to infinity) while for $0<n<4$ the singularity time is finite, but grows without bound  as $n$ gets closer to $4$ from below.} In other words, a sufficiently ``flat'' initial configuration suppresses the formation of singularities.}\\

\begin{table}
  \begin{center}
\def~{\hphantom{0}}
  \begin{tabular}{|l|c|c|c|}
      Case & $\Lambda_1(t)$ & $\Lambda_2(t)$ & $T_*$ \\[8pt]
      $n_- = +\infty,\; n_+ > 0$ $(r_- \neq 0)$ & $-\dfrac{r_-^2}{2{\gminus}}(S_* - S)^{-1}$ & $-\dfrac{r_-^2}{2{\gminus}}(S_* - S)^{-2}$ & $\infty$ \\[8pt]
      $n_- > 0,\; n_+ = +\infty$ $(r_- \neq R)$ & $\dfrac{r_-^2 - R^2}{2{\gminus}}(S_* - S)^{-1}$ & $\dfrac{r_-^2 - R^2}{2{\gminus}}(S_* - S)^{-2}$ & $\infty$ \\[8pt]
      $n_\pm = +\infty$ $(r_- \in [0,R])$ & $-\dfrac{R^2}{2{\gminus}}(S_* - S)^{-1}$ & $-\dfrac{R^2}{2{\gminus}}(S_* - S)^{-2}$ & $\infty$ \\
  \end{tabular}
  \caption{Asymptotic behaviors of the vorticity coordinate in the case of flat minimum ($n_- n_+ = \infty$), where $\Lambda_1(t)$ and $\Lambda_2(t)$ diverge as $t \to T_*$.}
  \label{table2}
  \end{center}
\end{table}

 {\noindent \textbf{b) Central minimum.}} Secondly, we consider the case of \textbf{central minimum}, where $r_- = 0$ and $n_{+} > 0$ , corresponding to a flow that becomes singular at the axis of symmetry with the stretching rate exhibiting a local minimum at the minimum point $r_0 = r_-$. These asymptotic behaviors are summarized in table \ref{table:3}. A critical threshold emerges at $n_+ = 4$, which we refer to as the \textbf{blowup threshold}. This means that the 3D Euler fluid under Lagrangian description remains regular if the local power-law approximation of the initial stretching rate satisfies {$n_+ \geq 4$}, but develops a finite-time singularity when {$n_+ < 4$}.

Since {larger} values of $n$ correspond to flatter profiles of $\gamma_0(r_0)$, we conclude that a qualitatively flat region around the minimum point significantly reduces the likelihood of singularity formation within finite time.

Another notable feature is the asymptotic behavior of the Lagrangian vorticity coordinate $(\omega, \gamma)$ and the vertical pathline component $z$ at the blowup point when the singularity time $T_*$ is finite. We divide the discussion into three main aspects:

\begin{itemize}
    \item[b.1)] The structure and finiteness of the singularity time $T_*$;
    \item[b.2)] The evolution of the plane vorticity $\omega$ and the vertical position $z$;
    \item[b.3)] The growth rate of the stretching rate $\gamma$, which determines the nature of the blowup.
\end{itemize}

These aspects provide a systematic framework for analyzing how the local geometry of the initial stretching rate $\gamma_0(r_0)$ influences the global regularity or singularity of the solution.\\

\begin{itemize}
    \item[b.1)] \textbf{Central Minimum: Singularity Time and Its Scaling Behavior}

    As shown in table \ref{table:3}, for $n \in (0, 2)$, the singularity time $T_*$ is given by:
\begin{equation}
    T_* = \frac{4\Lambda_{1*}^2}{|{\gminus}| R^4},
\end{equation}
where $\Lambda_{1*} = \lim_{t \to T_*} \Lambda_1(t)$ is a finite constant representing the limiting value of the integral $\Lambda_1(t)$ at the blowup point. Substituting the definition of $\Lambda_{1*}$ from equation \eqref{inte12}, we obtain the explicit dependence on the initial stretching rate:
\begin{equation}
\label{finalsingular}
    T_* = \frac{4}{|{\gminus}| R^4} \left( \int_0^R \frac{{\gminus} r_0}{\gamma_0(r_0) - {\gminus}} \mathrm{d}r_0 \right)^2 = \left[\frac{1}{\lambda_n - \varepsilon(\eta)}\right]^2 \frac{4}{(2 - n)^2 |{\gminus}| R^{2n}},
\end{equation}
where $\eta \in (0, R)$ is a constant derived from the integral mean value theorem.

From this expression, it follows that:

\begin{itemize}
    \item[i)] $T_* \propto |{\gminus}|^{-1}$, consistent with the scaling law established earlier in equation \eqref{meanT}. This implies that increasing the magnitude of the minimum of $\gamma_0(r_0)$ leads to an earlier singularity;

    \item[ii)] $T_* \propto R^{-2n}$, meaning that a larger cylindrical domain results in a shorter blowup time, as expected due to the enhanced spatial development of vorticity.
\end{itemize}

For {$n \in [2, 4)$}, deriving an explicit formula for $T_*$ becomes significantly more involved. The asymptotic approximation of the integral $I_{pq}(n; x)$ used in equations \eqref{firstipq} and \eqref{ipq} retains only the dominant term, omitting many subleading $O(1)$ contributions. Consequently, the full expression for the singularity time in this regime would be lengthy and cumbersome. We therefore refrain from presenting it explicitly here;\\
    
    \item[b.2)] \textbf{Central Minimum: Asymptotic Behavior of Plane Vorticity ($\omega$) and Vertical Pathline ($z$)}

    We analyze the Lagrangian plane vorticity $\omega$ and vertical pathline $z$ together due to their similar functional forms in equation \eqref{sol}. According to table \ref{table:3}, the behavior of these quantities near the blowup point depends strongly on the exponent $n$:

\begin{itemize}
    \item[i)] For $n \in (0, 2)$, both $\omega$ and $z$ decay linearly with $(T_* - t)$;

    \item[ii)] For $n \in (2, 4)$, they follow a slower power-law decay:
  \begin{equation}
      \omega, z \propto \left(T_* - t\right)^{\frac{2}{4 - n}}.
  \end{equation}
\end{itemize}

This distinction indicates that:

\begin{itemize}
    \item At the blowup point $r_- = 0$, the plane vorticity $\omega$ and the vertical position $z$ tend to zero as $t \to T_*$;

    \item The convergence speed differs between the two regimes: for $n \in (0, 2)$, the decay is independent of $n$, while for $n \in (2, 4)$, the convergence becomes slower as $n$ increases within that interval;

    \item  {The dependence on the cylinder radius is given by $\omega, z \propto R^{\frac{8}{4 - n} - 2} = R^{\frac{2n}{4-n}}$, showing that, for $n \in (0,4)$, larger radii slow down the decay of these quantities to zero.}
\end{itemize} 

At the critical case $n = 2$, the decay no longer follows a simple power law. Instead, we observe:
\begin{equation}
    \omega, z \propto \sqrt{T_* - t} \cdot \mathrm{e}^{W_{-1}\left(k \sqrt{T_* - t}\right)},
\end{equation}
where $k$ is a constant. This decay is faster than linear but slower than any $(T_* - t)^{\frac{2}{4 - n}}$ for $n \in (2, 4)$, preserving the ordering of convergence speeds across values of $n$;\\
    
    \item[b.3)] \textbf{Central Minimum: Divergence Rate of the Stretching Rate $\gamma$}

    In contrast to the convergence behavior of $\omega$ and $z$, the stretching rate $\gamma$ diverges at the blowup point with the same leading-order scaling across the entire range $n \in (0, 4)$:
\begin{equation}
    \gamma(r(r_-,t),t) \propto -\frac{1}{T_* - t}.
\end{equation}
However, the coefficient of this divergence differs depending on whether $n \in (0, 2]$ or $n \in (2, 4)$, as seen in table \ref{table:3}. In particular, there is an extra prefactor $\frac{2}{4 - n} \geq 1$ for $n \in [2, 4)$, so the growth of $\gamma$ is relatively faster in the latter interval compared to the former. Thus, the blowup of $\gamma$ occurs more rapidly for $n \in (0, 2]$, indicating a stronger singularity.
    
\end{itemize}

Based on this analysis, we identify a second critical threshold at $n = 2$, which governs the transition in the asymptotic structure of the solution. We refer to this as the \textbf{blowup-rate threshold}, distinguishing different rates of convergence and divergence among the vorticity coordinate components. {Incidentally, the example shown in section} \ref{parabola} {for the parabolic initial stretching rate (with minimum at $r_-=0$) corresponds to this critical blowup-rate threshold $n=2$.}\\

\begin{table}
  \begin{center}
\def~{\hphantom{0}}
  \begin{tabular}{|l|c|c|c|c|}
      Quantity & $n_+ \in (0,2)$ & $\{2\}$ & $(2,4)$ & $[4,+\infty)$ \\[8pt]
      $T_*$ & $\dfrac{4\Lambda_{1*}^2}{|{\gminus}| R^4}$ & $O(1)$ & $O(1)$ & $\infty$ \\[14pt]
      $S_* - S$ & $\dfrac{R^4}{4\Lambda_{1*}^2}(T_* - t)$ & $\dfrac{\lambda_2}{{\gminus}}\mathrm{e}^{2W_{-1}\left(-\frac{1}{2}R^2\Delta_{2t^+}\right)}$ & $\dfrac{\lambda_n}{{\gminus}}K_{2^+}^{\frac{n}{4-n}}(T_* - t)^{\frac{n}{4-n}}$ & N/A \\[8pt]
      $\omega$ & $-{\gminus}\omega_0(r_-)(T_* - t)$ & $\omega_0(r_-)\Delta_{2t^+}\mathrm{e}^{W_{-1}\left(-\frac{1}{2}R^2\Delta_{2t}\right)}$ & $\dfrac{L_2}{R^2}K_{2^+}^{\frac{2}{4-n}}\omega_0(r_-)(T_* - t)^{\frac{2}{4-n}}$ & N/A \\[8pt]
      $\gamma$ & $-\dfrac{1}{T_* - t}$ & $-\dfrac{1}{T_* - t}$ & $-\dfrac{2}{4-n}\dfrac{1}{T_* - t}$ & N/A \\[8pt]
      $z$ & $|{\gminus}| z_0(r_-)(T_* - t)$ & $z_0(r_-)\Delta_{2t^+}\mathrm{e}^{W_{-1}\left(-\frac{1}{2}R^2\Delta_{2t}\right)}$ & $\dfrac{L_2}{R^2}K_{2^+}^{\frac{2}{4-n}}z_0(r_-)(T_* - t)^{\frac{2}{4-n}}$ & N/A \\[8pt]
  \end{tabular}
  \caption{Asymptotic behavior of vorticity coordinate for central minimum ($r_- = 0$), where $K_{2^{+}}$, $L_{2^{+}}$, and $\Delta_{2t^{+}}$ are defined in equation \eqref{kl}.}
  \label{table:3}
  \end{center}
\end{table}

\begin{table}
  \begin{center}
\def~{\hphantom{0}}
  \begin{tabular}{|l|c|c|c|c|}
      Quantity & $n_\pm \in (0,1)$ & $\{1\}$ & $(1,2)$ & $[2,+\infty)$ \\[8pt]
      $T_*$ & $\dfrac{4\Lambda_{1*}^2}{|{\gminus}|R^4}$ & $O(1)$ & $O(1)$ & $\infty$ \\[14pt]
      $S_* - S$ & $\dfrac{R^4}{4\Lambda_{1*}^2}(T_* - t)$ & $\dfrac{\lambda_2}{{\gminus}}\mathrm{e}^{2W_{-1}\left(-\frac{R^2}{4r_-{\delta}}\Delta_{1t^{\pm}}\right)}$ & $\dfrac{\lambda_n}{{\gminus}}K_{1^{\pm}}^{\frac{n}{2-n}}(T_* - t)^{\frac{n}{2-n}}$ & N/A \\[8pt]
      $\omega$ & $-{\gminus}\omega_0(r_-)(T_* - t)$ & $\dfrac{1}{4}\omega_0(r_-)\Delta_{1t^{\pm}}\mathrm{e}^{2W_{-1}\left(-\frac{R^2}{4r_-{\delta}}\Delta_{1t^{\pm}}\right)}$ & $\dfrac{L_1}{R^2}K_{1^{\pm}}^{\frac{1}{2-n}}\omega_0(r_-)(T_* - t)^{\frac{1}{2-n}}$ & N/A \\[8pt]
      $\gamma$ & $-\dfrac{1}{T_* - t}$ & $-\dfrac{1}{T_* - t}$ & $-\dfrac{1}{2-n}\dfrac{1}{T_* - t}$ & N/A \\[8pt]
      $z$ & $|{\gminus}|z_0(r_-)(T_* - t)$ & $\dfrac{1}{4}z_0(r_-)\Delta_{1t^{\pm}}\mathrm{e}^{2W_{-1}\left(-\frac{R^2}{4r_-{\delta}}\Delta_{1t^{\pm}}\right)}$ & $\dfrac{L_1}{R^2}K_{1^{\pm}}^{\frac{1}{2-n}}z_0(r_-)(T_* - t)^{\frac{1}{2-n}}$ & N/A \\
  \end{tabular}
  \caption{The asymptotic behaviors of vorticity coordinate for non-central minimum with $r_- \in (0, R]$, $n_{\pm}\in\mathbb{R}_+$, where $K_{1^{\pm}}$, $L_{1^{\pm}}$ and $\Delta_{{1t}^{\pm}}$ are defined in equation \eqref{kl}.}
  \label{table:4}
  \end{center}
\end{table}

 {\noindent \textbf{c) Non-central minimum.}} Thirdly, we consider the case of \textbf{non-central minimum}, where the minimum point of the initial stretching rate lies away from the axis of symmetry, i.e., $r_- \in (0, R]$, and $\gamma_0(r_0)$ is continuous near this point with positive exponents $n_{\pm} \in \mathbb{R}_+$. The asymptotic behavior of the vorticity coordinate $(\omega, \gamma)$ and vertical pathline component $z$ in this scenario is summarized in table \ref{table:4}.

Comparing these results with those obtained for central minimum (table \ref{table:3}), we observe several qualitative similarities (with quantitative differences) in the structure of singularities and critical thresholds:\\

\begin{itemize}
    \item[c.1)] The \textbf{non-central minimum's blowup threshold} is now at $n = 2$, indicating that:
    \begin{itemize}
        \item[i)] For $n \in [2, +\infty)$, the solution remains regular for all finite times;

        \item[ii)] For $n \in (0, 2)$, a finite-time singularity occurs.
    \end{itemize}
    ~\\
    \item[c.2)] The \textbf{non-central minimum's blowup-rate threshold} appears now at $n = 1$, distinguishing different blowup speeds within the singular regime:
    \begin{itemize}
        \item[i)] For $n \in (0, 1)$, the asymptotic behavior of the vorticity coordinate is universal across all minimum positions (central or non-central), regardless of the radial position $r_- \in [0, R]$;

        \item[ii)] For $n = 1$, logarithmic corrections emerge;

        \item[iii)] For $n \in (1, 2)$, the decay rates of $\omega$ and $z$ depend explicitly on $n$, with slower convergence as $n$ approaches 2.
    \end{itemize}
\end{itemize}
~\\
These findings confirm that the local structure of the initial stretching rate $\gamma_0(r_0)$ near its minimum plays a dominant role in determining whether and how the    {Gibbon-Fokas-Doering model} develops a finite-time singularity. The critical thresholds $n = 1$ (blowup-rate threshold) and $n = 2$ (blowup threshold) define distinct regimes of behavior:

\begin{itemize}
    \item \textbf{Non-central minimum's no finite-time blowup:} $n \geq 2$;

    \item \textbf{Non-central minimum's finite-time blowup with universal scaling:} $n < 1$;

    \item \textbf{Non-central minimum's finite-time blowup with geometry-dependent scaling:} $1 < n < 2$.
\end{itemize}

{Incidentally, the example shown in section} \ref{parabola} {for the parabolic initial stretching rate (with minimum at $r_-=R$) corresponds to the critical case of blowup-rate threshold $n=1$.}\\

 {\noindent \textbf{Summary of key features across central and non-central blowups.}} {Now, we collect} the following key observations regarding the relative strength of blowups, {regardless of whether the blowup is central or non-central:}

\begin{itemize}
    \item For $n \in (0, 1)$, all non-flat scenarios exhibit identical blowup asymptotics, independent of the location of the blowup point (central or non-central);

    \item For $n \in [1, 2)$, the blowup for central minimum (i.e., at $r_- = 0$) develops more rapidly than the blowup for non-central minimum (i.e., at $r_- \in (0, R]$);

    \item For $n \in [2, 4)$, blowup  occurs only when the minimum point lies at the center; otherwise, for a non-central minimum, the solution remains regular;

    \item For $n \in [4, +\infty)$, no finite-time blowup occurs, irrespective of the location $r_-$ of the minimum of the vorticity stretching rate.
\end{itemize}

Thus, the closer the blowup point is to the cylinder's axis of symmetry, the faster the development of the singularity just before the blowup time.

The coefficient $K_{1^{\pm}}$, which appears in the asymptotic forms of $\omega$ and $z$, introduces an explicit dependence on $r_-$, the radial position of the blowup point. Since $r_-$ appears in the denominator of $K_{1^{\pm}}$, it follows that blowup points closer to the axis result in faster convergence of $\omega$ and $z$ to zero. However, this spatial sensitivity does not extend to the divergence rate of $\gamma$, which retains the universal scaling:
\begin{equation}
    \gamma \propto -(T_* - t)^{-1},
\end{equation}
regardless of whether the blowup occurs at the center or away from it.

In addition to $r_-$, another parameter influencing the asymptotic behavior is ${\delta}$, defined in equation \eqref{delta}, which quantifies the asymmetry between the left and right exponents $n_-$ and $n_+$ of $\gamma_0(r_0)$ near its minimum. Since ${\delta}$ also appears in the denominator of $K_{1^{\pm}}$, it affects the blowup dynamics in the interval $n \in (1, 2)$. Specifically, flows with asymmetric initial data—where $n_- \neq n_+$—exhibit faster blowup than those with symmetric profiles ($n_- = n_+$). In other words, \textit{a non-smooth or asymmetric profile of the initial stretching rate at the minimum point accelerates the formation of the singularity}.

{We emphasize a crucial physical distinction between central and non-central minimum scenarios. While the central minimum occurs on a point, the non-central minimum occurs on a ring, namely a degenerate case when compared to the point case. The ring case, due to the axisymmetry, has a perfectly ``flat'' profile in the azimuthal direction, which therefore changes the asymptotics in favor of more regular behavior, hence the required parameter $n$ for regular behavior is smaller compared with the central minimum case.} 

{Finally we briefly address the question: What happens when the location of the minimum of the initial stretching rate is shared by several rings, or by a combination of rings and the central point? Namely, when the set $\{x \in \Omega \quad |\quad  \gamma_0(x) = \min \gamma_0\}$ is the union of two or more elementary sets of the form $\{x \in \Omega \quad |\quad |x|=r_-\}$, with $r_- \geq 0$.  In such a case, the dominant ring (or center point) is the one for which the integral $\Lambda_1$ grows faster asymptotically as $t\to T_*$. Thus, for example, if the minima of $\gamma_0$ were at a ring with $n=2$ and at the center with $n=2$, then the asymptotic behavior of the system would be dominated by the ring with $n=2$ because the integral $\Lambda_1$ would diverge due to the stretching rate's profile at the ring, and thus the singularity time would still be $T_*=\infty$,  {even though a solitary initial minimum at the center with $n=2$ would lead to a finite-time singularity}. As another example, a ring with $n=1$ combined with a center point with $n=4$ would lead to no finite-time singularity because the integral $\Lambda_1$ would diverge due to the stretching rate's profile at the center,  {even though a solitary initial minimum at the ring with $n=1$ would lead to a finite-time singularity.}}

\section{Conclusion}\label{conclusion}

We have applied the vorticity-stretching ansatz by \cite{gibbon1999dynamically} on cylindrical domains with boundaries, to explore the potential formation of singularities in the three-dimensional incompressible Euler equations. The study incorporates axisymmetry, a cylindrical spatial domain, and no-flow boundary conditions. By employing the Lagrangian framework, explicit solutions for the planar vorticity, its stretching rate, and fluid pathlines are obtained through the use of conserved quantities—constants of motion—that remain invariant along particle trajectories. A key new result stemming from our axisymmetry assumption is that explicit solutions for the Eulerian velocity components can be obtained  {under specific initial stretching rate, such as the parabolic initial $\gamma_0(r_0)$ discussed in Section \ref{parabola}}.

The analysis reveals that the key factor determining whether or not a singularity forms in finite time is the local behavior of the initial vorticity stretching rate near the position of its minimum. Specifically, the  {position} of this minimum (whether it is  at the center point or at a ring) and the way the  {initial} vorticity stretching rate behaves in its vicinity dictate the development of blowup in the solution. These are considered local properties of the initial data and do not depend on the global configuration of the flow.  {An exception to this result occurs when the minimum of $\gamma$ is initially attained at several places. Then, the structure of the minimum at each of these places needs to be considered in order to determine which one dominates, based on the asymptotic behavior of the corresponding auxiliary integrals $\Lambda_1$ defined in equation \eqref{lam1}.}

Further investigation into the asymptotic behavior of the system shows that different power-law approximations of the initial vorticity stretching rate near the position of its minimum lead to distinct types of singularity. We identified two critical thresholds: one marking the transition from regular to singular solutions (in terms of the formula for the singularity time $T_*$), and another one distinguishing different rates of  {blowup or vanishing} near the singularity time. {When the minimum is located at the center ($r_0=0$), no finite-time singularity occurs if the exponent $n$ of the local power-law approximation satisfies $n \geq 4$ (e.g., $\gamma_0(r_0) = 3r_0^4 - 1$ where $n=4$), whereas a finite-time singularity develops for $0 < n < 4$ (e.g., $\gamma_0(r_0) = 2r_0^2 - 1$ where $n=2$). Conversely, when the minimum lies off-center at $r_0 > 0$, the critical threshold shifts to $n \geq 2$ for the absence of a finite-time singularity (e.g., $\gamma_0(r_0) = 40(r_0-0.5)^{3} - 1$ where $n=3$), while $0 < n < 2$ leads to finite-time singularity formation (e.g., $\gamma_0(r_0) = 5\sqrt{2}|r_0-0.5|^{\frac{3}{2}} - 1$ where $n=\frac{3}{2}$). The detailed dependence of the singularity time on the local behavior of $\gamma_0(r_0)$ at its minimum is summarized in Tables \ref{table2}, \ref{table:3} and \ref{table:4}.}

The emergence of a ring-shaped blowup region at the ``corner'' boundary of a cylinder brings to mind the numerical observations in  \cite{luo2014potentially}, where a ring-localized singularity was detected in the 3D axisymmetric Euler equations. Our theoretical framework, although in a different class of solutions with respect to the reflection $z\to -z$ (in our case both $u_r$ and $u_\theta$ are even functions of $z$, while in Luo and Hou only $u_r$ is even in $z$, while $u_\theta$ is \emph{odd} in $z$), allows for ring-shaped minimum profiles of $\gamma_0(r_0)$, provides analytical support for such non-central singularities and explains the lower blowup threshold in this configuration due to geometric inhibition. 

Interestingly, there is a work in the literature that belongs to the class studied in our paper, and therefore can be discussed quantitatively in terms of our results: the paper by Ohkitani and Gibbon \cite[Section II]{ohkitani2000numerical}{, where two ring-like solutions were computed numerically: \emph{Initial condition 2} (in Section II B) and \emph{Initial condition 3} (in Section II C). \emph{Initial condition 2} corresponds to a non-central minimum with $n = 2$, while \emph{Initial condition 3} corresponds to a central minimum with $n = 2$. According to our theoretical framework, for \emph{Initial condition 2} the exponent $n$ lies at the blowup threshold (hence regular behavior is expected), while for \emph{Initial condition 3} the exponent $n$ lies in the blowup-rate threshold (hence singular behavior is expected). This prediction is in agreement with the numerical results reported in their study: compare figure 8 (for \emph{Initial condition 2}), showing a tame behavior, with figure 11 (for \emph{Initial condition 3}), showing a singular behavior. This consistency between our theoretical predictions and prior numerical simulations underscores the validity and predictive power of the proposed framework in characterizing singularity formation in such systems.}

Our work not only provides a platform for studying the fundamental question of regularity in three-dimensional Euler flows but also serves as a mathematical illustration of vortex interactions under {Gibbon-Fokas-Doering} {vorticity-stretching} model. It highlights how local geometric features of the initial conditions can drive the formation of singularities, offering insight into the mechanisms underlying turbulent behavior in idealized flows.

\section{Acknowledgements}
This publication has emanated from research supported in part by a grant from Taighde \'Eireann – Research Ireland under Grant number 18/CRT/6049. For the purpose of Open Access, the authors have applied a CC BY public copyright licence to any Author Accepted Manuscript version arising from this submission. AI tools were used for translation and language editing; all scientific content, analysis, and
conclusions were produced by the authors.

\bibliographystyle{jfm}
\bibliography{singularity_arxiv}

@article{euler1761principia,
  title={Principia motus fluidorum},
  author={Euler, Leonhard},
  journal={Novi commentarii academiae scientiarum Petropolitanae},
  pages={271--311},
  year={1761}
}

@article{beale1984remarks,
  title={Remarks on the breakdown of smooth solutions for the 3-D Euler equations},
  author={Beale, Thomas J and Kato, Tosio and Majda, Andrew},
  journal={Communications in Mathematical Physics},
  volume={94},
  number={1},
  pages={61--66},
  year={1984},
  publisher={Springer}
}

@article{constantin1996geometric,
  title={Geometric constraints on potentially singular solutions for the 3-D Euler equations},
  author={Constantin, Peter and Fefferman, Charles and Majda, Andrew J},
  journal={Communications in Partial Differential Equations},
  volume={21},
  number={3-4},
  pages={559--571},
  year={1996}
}

@article{drivas2023singularity,
  title={Singularity formation in the incompressible Euler equation in finite and infinite time},
  author={Drivas, Theodore D and Elgindi, Tarek M},
  journal={EMS Surveys in Mathematical Sciences},
  volume={10},
  number={1},
  pages={1--100},
  year={2023}
}

@article{gibbon1999dynamically,
  title={Dynamically stretched vortices as solutions of the 3D Navier--Stokes equations},
  author={Gibbon, John D and Fokas, Athanassios S and Doering, Charles R},
  journal={Physica D: Nonlinear Phenomena},
  volume={132},
  number={4},
  pages={497--510},
  year={1999},
  publisher={Elsevier}
}

@article{elgindi2021incompressible,
  title={The incompressible Euler equations under octahedral symmetry: singularity formation in a fundamental domain},
  author={Elgindi, Tarek M and Jeong, In-Jee},
  journal={Advances in Mathematics},
  volume={393},
  pages={108091},
  year={2021},
  publisher={Elsevier}
}

@article{elgindi2020finite,
  title={Finite-time singularity formation for strong solutions to the Boussinesq system},
  author={Elgindi, Tarek M and Jeong, In-Jee},
  journal={Annals of PDE},
  volume={6},
  number={1},
  pages={5},
  year={2020},
  publisher={Springer}
}

@article{martinez2024evidence,
  title={Evidence of a finite-time pointlike singularity solution for the Euler equations for perfect fluids},
  author={Mart{\'\i}nez-Arg{\"u}ello, Diego and Rica, Sergio},
  journal={Physical Review Fluids},
  volume={9},
  number={9},
  pages={094401},
  year={2024},
  publisher={APS}
}

@article{lundgren1982strained,
  title={Strained spiral vortex model for turbulent fine structure},
  author={Lundgren, Thomas S},
  journal={Physics of Fluids},
  volume={25},
  number={12},
  pages={2193--2203},
  year={1982}
}

@article{childress1989blow,
  title={Blow-up of unsteady two-dimensional Euler and Navier-Stokes solutions having stagnation-point form},
  author={Childress, Stephen and Ierley, Glenn R and Spiegel, Edward A and Young, William R},
  journal={Journal of Fluid Mechanics},
  volume={203},
  pages={1--22},
  year={1989},
  publisher={Cambridge University Press}
}

@article{choi2017finite,
  title={On the Finite-Time Blowup of a One-Dimensional Model for the Three-Dimensional Axisymmetric Euler Equations},
  author={Choi, Kyudong and Hou, Thomas Y and Kiselev, Alexander and Luo, Guo and Sverak, Vladimir and Yao, Yao},
  journal={Communications on Pure and Applied Mathematics},
  volume={70},
  number={11},
  pages={2218--2243},
  year={2017},
  publisher={Wiley Online Library}
}

@article{gibbon2008three,
  title={The three-dimensional Euler equations: Where do we stand?},
  author={Gibbon, John D},
  journal={Physica D: Nonlinear Phenomena},
  volume={237},
  number={14-17},
  pages={1894--1904},
  year={2008},
  publisher={Elsevier}
}

@article{luo2014potentially,
  title={Potentially singular solutions of the 3D axisymmetric Euler equations},
  author={Luo, Guo and Hou, Thomas Y},
  journal={Proceedings of the National Academy of Sciences},
  volume={111},
  number={36},
  pages={12968--12973},
  year={2014},
  publisher={National Acad Sciences}
}

@article{hou2023potentially,
  title={Potentially singular behavior of the 3D Navier--Stokes equations},
  author={Hou, Thomas Y},
  journal={Foundations of Computational Mathematics},
  volume={23},
  number={6},
  pages={2251--2299},
  year={2023},
  publisher={Springer}
}

@article{constantin2000euler,
  title={The Euler equations and non-local conservative Riccati equations},
  author={Constantin, Peter},
  journal={International Mathematics Research Notices},
  year={2000},
  volume={9},
  pages={455--465}
}

@article{ohkitani2000numerical,
  title={Numerical study of singularity formation in a class of Euler and Navier--Stokes flows},
  author={Ohkitani, Koji and Gibbon, John D},
  journal={Physics of Fluids},
  volume={12},
  number={12},
  pages={3181--3194},
  year={2000},
  publisher={AIP Publishing}
}

@article{larios2018computational,
  title={A computational investigation of the finite-time blow-up of the 3D incompressible Euler equations based on the Voigt regularization},
  author={Larios, Adam and Petersen, Mark R and Titi, Edriss S and Wingate, Beth},
  journal={Theoretical and Computational Fluid Dynamics},
  volume={32},
  pages={23--34},
  year={2018},
  publisher={Springer}
}

@article{stuart1988Nonlinear,
  title={Nonlinear Euler Partial Differential Equations: Singularities in their Solution},
  author={ Stuart, Trevor J },
  journal={Applied Mathematics Fluid Mechanics Astrophysics World Scientific},
  volume={-1},
  pages={81-95},
  year={1988},
}

@article{gibbon2003exact,
  title={Exact, infinite energy, blow-up solutions of the three-dimensional Euler equations},
  author={Gibbon, John D and Moore, Daniel R and Stuart, Trevor J},
  journal={Nonlinearity},
  volume={16},
  number={5},
  pages={1823},
  year={2003},
  publisher={IOP Publishing}
}

@article{bustamante2022role,
  title={On the role of continuous symmetries in the solution of the three-dimensional Euler fluid equations and related models},
  author={Bustamante, Miguel D},
  journal={Philosophical Transactions of the Royal Society A},
  volume={380},
  number={2226},
  pages={20210050},
  year={2022},
  publisher={The Royal Society}
}

@article{mulungye2016atypical,
  title={Atypical late-time singular regimes accurately diagnosed in stagnation-point-type solutions of 3D Euler flows},
  author={Mulungye, Rachel M and Lucas, Dan and Bustamante, Miguel D},
  journal={Journal of Fluid Mechanics},
  volume={788},
  pages={R3},
  year={2016},
  publisher={Cambridge University Press}
}

@article{mulungye2015symmetry,
  title={Symmetry-plane model of 3D Euler flows and mapping to regular systems to improve blowup assessment using numerical and analytical solutions},
  author={Mulungye, Rachel M and Lucas, Dan and Bustamante, Miguel D},
  journal={Journal of Fluid Mechanics},
  volume={771},
  pages={468--502},
  year={2015},
  publisher={Cambridge University Press}
}

@article{kukavica2011analyticity,
  title={On the analyticity and Gevrey-class regularity up to the boundary for the Euler equations},
  author={Kukavica, Igor and Vicol, Vlad},
  journal={Nonlinearity},
  volume={24},
  number={3},
  pages={765},
  year={2011},
  publisher={IOP Publishing}
}

@book{rodino1993linear,
  title={Linear partial differential operators in Gevrey spaces},
  author={Rodino, Luigi},
  year={1993},
  publisher={World Scientific}
}

@book{batchelor2000introduction,
  title={An introduction to fluid dynamics},
  author={Batchelor, George Keith},
  year={2000},
  publisher={Cambridge University Press}
}

\end{document}